\documentclass[10pt,aps,prd,twocolumn,nofootinbib,]{revtex4-1}

\usepackage[pdftex]{graphicx}
\usepackage{latexsym,amsmath,amssymb,lmodern,float,url,bm,stmaryrd}
\usepackage{enumerate}
\usepackage{ulem}
\usepackage{braket,feynmf,slashed}
\usepackage{bbold}
\usepackage{appendix} 
\usepackage{natbib}
\usepackage{xspace}
\usepackage{tikz-feynman}
\usepackage[pdftex,bookmarks,linktocpage,pdfpagelabels,plainpages=false,hyperfigures,linkcolor=blue,citecolor=blue]{hyperref} 
\hypersetup{colorlinks=true}

\newcommand\ee{\end{equation}}
\newcommand\be{\begin{equation}}
\newcommand\eea{\end{eqnarray}}
\newcommand\bea{\begin{eqnarray}}
\newcommand\cns{${\rm{CE}}\nu{\rm{NS}}$\xspace}
\newcommand\nn{ \nonumber \\}
\newcommand\mr{\mathrm}
\newcommand\mc{\mathcal}
\newcommand\twotwo{{(2\shortrightarrow 2)}}
\newcommand\twothree{{(2\shortrightarrow 3)}}

\newcommand\vk{|\vec{k}|}

\newcommand\vkp{|\vec{k'}|}

\newcommand\mchi{m_{\chi}}
\newcommand\vcl{\vec{v}_{\chi,{\rm{Lab}}}}
\newcommand\vcm{\vec{v}_{\chi,{\rm{CoM}}}}
\newcommand\vtm{\vec{v}_{T,{\rm{CoM}}}}

\begin{document}
\bibliographystyle{apsrev4-1}
\title{The Migdal Effect and Photon Bremsstrahlung in effective field theories of dark matter direct detection and coherent elastic neutrino-nucleus scattering}
\author{Nicole F.~Bell$^{\bf a}$, James B.~Dent$^{\bf b,c}$, Jayden L.~Newstead$^{\bf d}$, Subir Sabharwal$^{\bf e}$, Thomas J.~Weiler$^{\bf f}$}

\affiliation{$^{\bf a}$ ARC Centre of Excellence for Particle Physics at the Terascale \\
School of Physics, The University of Melbourne, Victoria 3010, Australia}
\affiliation{$^{\bf b}$ Department of Physics, Sam Houston State University, Huntsville, TX 77341, USA}
\affiliation{$^{\bf c}$ Kavli Institute for Theoretical Physics,
University of California, Santa Barbara, CA 93106-4030, USA}
\affiliation{$^{\bf d}$ Department of Physics, Arizona State University, Tempe, AZ 85287, USA}
\affiliation{$^{\bf e}$ Eastmore Group, Suite 1701, 40 Wall St, New York, NY 10005, USA}
\affiliation{$^{\bf f}$ Department of Physics and Astronomy, Vanderbilt University, Nashville, TN 37235, USA}

\begin{abstract}
Dark matter direct detection experiments have limited sensitivity to light dark matter (below a few GeV), due to the challenges of lowering energy thresholds for the detection of nuclear recoil to below $\mc{O}(\mr{keV})$. While impressive progress has been made on this front, light dark matter remains the least constrained region of dark-matter parameter space. It has been shown that both ionization and excitation due to the Migdal effect and coherently-emitted photon bremsstrahlung from the recoiling atom can provide observable channels for light dark matter that would otherwise have been missed owing to the resulting nuclear recoil falling below the detector threshold. In this paper we extend previous work by calculating the Migdal effect and photon bremmstrahlung rates for a general set of interaction types, including those that are momentum-independent or -dependent, spin-independent or -dependent, as well as examining the rates for a variety of target materials, allowing us to place new experimental limits on some of these interaction types. Additionally, we include a calculation of these effects induced by the coherent scattering on nuclei of solar or atmospheric neutrinos. We demonstrate that the Migdal effect dominates over the bremsstrahlung effect for all targets considered for interactions induced by either dark matter or neutrinos. This reduces photon bremsstrahlung to irrelevancy for future direct detection experiments. 

\end{abstract}

\maketitle

\section{Introduction}
Although dark matter is the most abundant form of matter in the universe, its particle nature, including its mass and non-gravitational interactions with Standard Model particles, remains elusive. For dark matter masses $m_\chi \gtrsim 1~\rm{GeV}$, searches for dark matter induced nuclear recoils in detectors with extremely low backgrounds are a primary means of attempting to discern the particle nature of dark matter ~\cite{Aprile:2018dbl,Agnese:2018col,Crisler:2018gci,Ren:2018gyx,Agnes:2018ves,Akerib:2017kat,Akerib:2016vxi,Akimov:2017ade,Agnes:2017grb,Agnese:2017njq,Agnese:2017jvy,Angloher:2017sxg,Petricca:2017zdp,Aguilar-Arevalo:2016ndq,Amole:2017dex}. Although such direct detection experiments continue to increase their sensitivity to possible dark matter interactions by lowering their energy thresholds, increasing their exposure, and reducing backgrounds, no conclusive signals of such interactions have arisen. This situation has spurred a great deal of interest in examining alternative signals to nuclear recoils.  For example, dark matter could produce ionization signals through alternate channels. These include dark matter scattering directly from electrons ~\cite{Essig:2011nj,Essig:2012yx,Hochberg:2015pha,Lee:2015qva,Essig:2015cda,Roberts:2016xfw,Emken:2017erx,Essig:2017kqs,Agnes:2018oej,Crisler:2018gci,Agnese:2018col}, or secondary signals such as the Migdal effect~\cite{Ibe:2017yqa,Dolan:2017xbu,Akerib:2018hck,Armengaud:2019kfj} and photon bremsstrahlung~\cite{Kouvaris:2016afs,McCabe:2017rln}. Such effects are generically sub-dominant to the conventional nuclear scattering, with the exception of very light dark matter with a mass below a few GeV. In such cases these alternate signals can provide competitive experimental sensitivities. In the present work we examine both the Migdal effect and photon bremsstrahlung from the nucleus, in cases where the accompanying nuclear recoil is manifested below detector thresholds. 

We extend previous analyses by formulating both the dark matter induced Migdal effect and photon bremsstrahlung in an effective field theory (EFT) framework of nucleons ~\cite{Fan:2010gt,Fitzpatrick:2012ix,Fitzpatrick:2012ib,Anand:2013yka,Dent:2015zpa,Baum:2017kfa}. We examine a variety of target materials and interaction types, and demonstrate that the scattering rates due to the bremsstrahlung effect are sub-dominant to those from the Migdal effect in all cases. We then proceed to place new bounds on EFT operators for low mass dark matter.

It is well known that coherent elastic neutrino-nucleus scattering (\cns) from solar and atmospheric sources can approximate WIMP-style interactions, and will eventually become a background for upcoming direct detection experiments \cite{Billard:2013qya}. Therefore, we also provide the expected rates for these sources due to both the Migdal and bremsstrahlung processes. We find that there exists a small window in recoil energy where the Migdal effect from solar neutrinos can provide comparable rates to those from atmospheric neutrino induced nuclear recoils. This provides the possibility of an increased background through the means of \cns. However, these effects are below the standard nuclear recoil rates in the remainder of the recoil energy space outside this small window.

This work proceeds as follows. In Sec.~\ref{sec:migdal_brem} we briefly review both the Migdal effect and photon bremsstrahlung in the direct detection context, followed by an overview in Sec.~\ref{sec:generalEFT} of the general effective field theory framework for dark matter-nucleus scattering. In Sec.~\ref{sec:rates} we demonstrate the dominance of the Migdal effect across all interactions and materials. In Sec. \ref{sec:limits} we use the Migdal effect rates to place new bounds on EFT operators using existing direct detection data.  The results of including the Migdal effect and bremsstrahlung in coherent elastic neutrino-nucleus scattering arising from the scattering of solar and atmospheric neutrinos off of target nuclei are given in Sec.~\ref{sec:neutrino}. A summary of our results and directions for future work are provided in Sec.~\ref{sec:summary}.


\section{Migdal Effect and photon bremsstrahlung in dark matter-nucleus scattering}
\label{sec:migdal_brem}

\subsection{General recoil rate for dark matter scattering including the Migdal effect}

The Migdal effect is the process of ionization or excitation of an atom due to the lack of instantaneous movement of the electron cloud during a nuclear recoil event, leading to a possible detection of the subsequent electromagnetic signature.  We will briefly review the treatment of \cite{Ibe:2017yqa} in order to present the formulae needed in the present work to calculate rates for the Migdal effect due to both dark matter scattering and \cns.

To calculate the rate of ionization events due to the Migdal effect the standard dark matter-nucleus differential recoil rate, $R_{\chi T}$ (where $T$ denotes the target nucleus), is multiplied by the ionization rate, $|Z_{{\rm{ion}}}|^2$, 
\bea
\label{eq:migdal_rate}
\frac{d^2R}{dE_Rdv} &=& \frac{d^2R_{\chi T}}{dE_Rdv} \times |Z_{{\rm{ion}}}|^2.
\eea
In what follows $E_R$ is the nuclear recoil energy, and $v$ is the dark matter speed,\footnote{Consistent with the treatment in \cite{Ibe:2017yqa}, we do not include the effects of electronic excitation because the excitation probabilities are suppressed in comparison to those from ionization.}
The ionization rate corresponding to single electron ionization, $|Z_{\rm{ion}}|^2$ is given in terms of the ionization probability $p_{q_e}^c$ (here the subscript $q_e$ refers to the average momentum transfer to an individual electron in the target) and electronic energy, $E_e$, by
\bea
|Z_{{\rm{ion}}}|^2 = \frac{1}{2\pi}\sum_{n,\ell}\int dE_e\frac{d}{dE_e}p^c_{q_e}(n\ell\rightarrow(E_e)).
\eea
Therefore, one finds the differential rate
\bea
\label{eq:migdal_triple}
\frac{d^3R}{dE_RdE_{{\rm{EM}}}dv} = \frac{d^2R_{\chi T}}{dE_Rdv} \times\frac{1}{2\pi}\sum_{n,\ell}\frac{d}{dE_e}p^c_{q_e}(n\ell\rightarrow(E_e))\nn
\eea
where the total electromagnetic energy injection $E_{{\rm EM}}$ is defined as the sum of the outgoing unbounded electron energy, $E_e$ and energy from de-excitation $E_{n\ell}$. The final atomic state is assumed to be completely de-excited, and so $E_{n\ell}$ is taken to be the binding energy of the $(n,\ell)$ state from which the outgoing electron was ejected. The differential probability rates are calculated in \cite{Ibe:2017yqa} with the use of the publicly available {{\tt Flexible Atomic Code (FAC, cFAC)}} \cite{Gu:2008}\footnote{In this work we utilize the probabilities calculated by the authors of \cite{Ibe:2017yqa}, who graciously made the numerical results presented in their Fig.~4 available to us.}. This treatment makes no assumption about the underlying nuclear scattering interaction, and is therefore applicable to the general EFT interaction framework to be introduced in Sec.~\ref{sec:generalEFT}

It should be noted that this formulation of the Migdal effect treats each atom in isolation, not taking into account the shifts in electronic energy levels due to atoms in a liquid (such as xenon and argon targets) or a crystal (such as sodium iodide, germanium, or silicon targets). This prompted the EDELWEISS collaboration, for example, to investigate the Migdal effect with respect to the $n = 3$ level electrons exclusively, rather than including the valence $n = 4$ shell \cite{Armengaud:2019kfj}. In this work, we also adopt the free atom approximation, though clearly a more complete formulation that integrates the effects of atoms in the detector environment would be welcomed.

This approach to the inclusion of ionization via the Migdal effect is not exclusive to scattering induced by dark matter. Nuclear recoils produced by \cns can be treated in a similar fashion (as noted in Sec.VII of \cite{Ibe:2017yqa}), as we will discuss shortly. 

Before displaying the rates for the Migdal effect, we will next describe dark matter-nucleus scattering accompanied by photon bremsstrahlung.

\subsection{Photon bremsstrahlung and dark matter-nucleus scattering}

In \cite{Kouvaris:2016afs}, the authors showed that the irreducible contribution of the inelastic scattering process $\chi(p) + N(k) \rightarrow \chi(p') + N(k') + \gamma(\omega)$ which is associated with the elastic scattering $\chi(p) + N(k) \rightarrow \chi(p') + N(k')$, can provide a detectable signal for a range of dark matter masses which would not produce elastic nuclear recoils above the detector threshold, thereby enhancing the sensitivity of direct detection experiments to low mass dark matter.

The double differential cross section in nuclear recoil energy, $E_R$, and photon energy, $\omega$, is found by integrating the bremsstrahlung matrix element amplitude over the 3-body phase space (details of this integration are given in Appendix~\ref{app:phaseSpace}). In Ref.~\cite{Kouvaris:2016afs}, the bremsstrahlung cross section was found in terms of the factorized elastic 2-to-2 cross section,
\bea
\label{eq:PradlerBrem}
\frac{d^2\sigma}{dE_Rd\omega} 
&=& \frac{4\alpha Z^2}{3\pi}\frac{E_R}{m_T \omega}\left(\frac{d\sigma}{dE_R}\right)_\twotwo,
\eea
where $m_T$ is the target nuclei mass (note Ref.~\cite{Kouvaris:2016afs} uses $m_N$ for the nuclear mass, while we define $m_N$ as the nucleon mass). This result is obtained in the soft photon limit where $\omega/m_\chi \ll 1$. The bremsstrahlung rate can then be cast into the same form as Eq.~(\ref{eq:migdal_triple}),
\bea
\label{eq:PradlerBremRate}
\frac{d^3 R}{dE_Rd\omega dv} 
&=& \frac{d^2R_{\chi T}}{dE_R dv}\frac{4\alpha Z^2}{3\pi}\frac{E_R}{m_T \omega}.
\eea
The elastic scattering process considered in \cite{Kouvaris:2016afs} was momentum-independent (here and hereafter meaning, no dependence on the transferred momentum) as well as independent of the nuclear spin. Thus, the cross section benefited from a coherent enhancement factor $A^2$, where $A$ is the number of protons plus neutrons. In general, however, the elastic scattering process may proceed through momentum-dependent or spin-dependent interactions. Importantly, the key relations for both the Migdal effect and bremsstrahlung processes are independent of the dark matter-nucleon interaction. The bremsstrahlung rate given in Eq.(\ref{eq:PradlerBremRate}) does not account for atomic screening effects as detailed in Ref.~\cite{Kouvaris:2016afs}. As this reduces the bremsstrahlung rate at low energies, which we demonstrate to be sub-dominant to the Migdal effect, the screening effect has been neglected throughout the remainder of this work.

We will now turn to a discussion of the general non-relativistic effective field theory (NR-EFT) for dark matter-nucleus scattering. This NR-EFT provides the formalism within which we will calculate the elastic rate appearing in Eq.~(\ref{eq:migdal_triple}) and Eq.~(\ref{eq:PradlerBremRate}) for the Migdal effect and photon bremsstrahlung, respectively.


\section{Overview of Effective Field Theory of Dark Matter-Nucleus Scattering}
\label{sec:generalEFT}

Dark matter candidates can take a variety of forms, many of which do not couple to the standard model through the basic momentum-independent channels~\cite{Barger:2010gv,Chang:2009yt,Ho:2012bg}. Momentum-dependent interactions are suppressed relative to interactions with no momentum dependence since galactic dark matter is non-relativistic with a virial speed of roughly $v\sim\mathcal{O}(10^{-3})$. The target nucleus will recoil with energy $E_R$ due to a momentum transfer $\vec{q}$, with magnitude dependent on the target nuclear mass $m_T$, as $|\vec{q}| = \sqrt{2m_T E_R}$. The recoil energy is given by $E_{R} = \mu_T^2 v^2(1-\cos\theta)/m_T$. Here $\theta$ is the scattering angle between the incident scatterer and the recoiling nucleus, and $\mu_T = m_T m_\chi/(m_T + m_\chi)$ is the dark matter-nucleus reduced mass. For typical weak scale masses $m_{\chi} = 100$ GeV $\approx m_T$, the recoil energy will be in the $\mathcal{O}(10{\rm{keV}})$ range. These values lead to momenta transfer of roughly $|\vec{q}| \lesssim 100~{\rm{MeV}}$, and therefore terms containing $\vec{q}/m_{\chi\,{\rm{or}}\, N}$, will be suppressed relative to terms without momentum dependence. Spin-dependent interactions do not contain the $A^2$ enhancement factor arising for isospin invariant interactions, but rather depend on the nuclear spin and thus on the expectation values of the spins of the nucleons. As a result, spin-dependent interactions are suppressed by several orders of magnitude compared to the spin-independent variety.

The NR-EFT formalism contains general interactions which can include momentum- and spin-dependence. This approach aids in mapping high energy particle models of dark matter onto operators that act on nuclear states~\cite{Fan:2010gt,Fitzpatrick:2012ix}. This formalism provides a neat split between the nuclear and particle physics, allowing us to use pre-calculated nuclear form factors for the relevant interaction operators. In this formalism a general set of operators are compiled from the relevant variables: the WIMP spin $\vec{S}_\chi$, nucleon spin $\vec{S}_N$, momentum transfer $\vec{q}$, and the projected velocity $\vec{v}^\perp$ (where $\vec{q}\cdot\vec{v}^\perp=0$). Here we consider a subset of the typical set of operators \cite{Fitzpatrick:2012ix},
\bea
\begin{array}{cc}
\mathcal{O}_1 & \mathbb{1}_\chi\mathbb{1}_N
\\
\mathcal{O}_4 & \vec{S}_\chi \cdot \vec{S}_N
\\
\mathcal{O}_6 & \left(\frac{\vec{q}}{m_N}\cdot\vec{S}_{\chi}\right)\left(\frac{\vec{q}}{m_N}\cdot\vec{S}_{N}\right)
\\
\mathcal{O}_{10} & \mathbb{1}_\chi \left(i\frac{\vec{q}}{m_N}\cdot\vec{S}_N\right)
\end{array}
\eea
Inclusion of just this subset of operators allows us to examine the effects of spin-independent/dependent and momentum-independent/dependent WIMP interactions, without needing to consider the full space of operators. For small momentum transfer (i.e., low mass WIMP scattering) under consideration here, these operators can be considered representative of the behavior of operators with the same momentum dependence \cite{Dent:2016iht}.

\subsection{From nucleons to target nuclei:  nuclear response functions}
Dark matter-nucleus scattering, and the resulting electronic ionization due to the Migdal effect or photon bremsstrahlung, is a coherent process (meaning the momentum transfer is less than the inverse nuclear size). This implies that the EFT nucleon operators in the amplitude must be summed over the nucleons in the target nuclei. For the bremsstrahlung process, incoherence caused by internucleon effects are neglected since it only makes a meaningful contribution at photon energies~\cite{Maydanyuk:2012yn,Maydanyuk:2013aqa} higher than those of interest here. 

Following the formalism of Fitzpatrick et al.~\cite{Fitzpatrick:2012ix}, the operators acting on nuclear states are decomposed into spherical components and expanded in multipoles. For this analysis, only three single-particle operators from semi-leptonic electroweak theory \cite{Donnelly:1978tz,Donnelly:1979ezn,Walecka:1995mi} are required: $M$, $\Sigma'$, $\Sigma''$, called the vector charge, axial transverse electric and axial longitudinal operators, respectively:
\begin{flalign}
M_{JM;\tau}(q^2) &\equiv \sum_{i=1}^A M_{JM}(q\vec{x}_i)\tau_3(i),\nn
\Sigma^{'}_{JM;\tau}(q^2) &\equiv -i\sum_{i=1}^A\left\{\frac{1}{q}\vec{\nabla}_i\times\vec{M}^M_{JJ}(q\vec{x}_i)\right\}\cdot\vec{\sigma}(i)\tau_3(i),\nn
\Sigma^{''}_{JM;\tau}(q^2) &\equiv \sum_{i=1}^A\left\{\frac{1}{q}\vec{\nabla}_i{M}_{JM}(q\vec{x}_i)\right\}\cdot\vec{\sigma}(i)\tau_3(i),
\end{flalign}
where $\tau$ is an isospin label. The spin summed elastic WIMP-nucleus scattering amplitude can then be written as
\begin{align}
\label{eq:Amp}
&\frac{1}{2j_\chi +1}\frac{1}{2j_N +1} 
 \sum_{\mr{spins}}\vert\mc{M}\vert^2 \nn 
& = \frac{4\pi}{2j_N+1}  \sum_{\tau=0,1}  \sum_{\tau'=0,1} \left[ R^{\tau\tau'}_{M}(q^2) W^{\tau\tau'}_{M}(q^2) \right.\nn
& \qquad\qquad + \left. R^{\tau\tau'}_{\Sigma'}(q^2)  W^{\tau\tau'}_{\Sigma'}(q^2) + R^{\tau\tau'}_{\Sigma''}(q^2)  W^{\tau\tau'}_{\Sigma''}(q^2)\right],
\end{align}
where $j_\chi$ and $j_N$ label the WIMP and nuclear spin, respectively. The momentum transfer q dependent functions $R_X(q^2)$ and $W_X(q^2)$ are known as the WIMP and nuclear response functions, respectively. Note that the definition of the amplitude $\mathcal{M}$ in Eq.~(\ref{eq:Amp}) has the non-relativistic normalization and produces a dimensionful amplitude (a factor of $4m_Tm_\chi$ is required to make it dimensionless). The WIMP response functions are
\bea
R^{\tau\tau'}_{M}(q^2) & = & c_1^\tau c_1^{\tau'},\nn
R^{\tau\tau'}_{\Sigma'}(q^2) & = & \frac{j_\chi(j_\chi+1)}{12}c_4^\tau c_4^{\tau'},\nn
R^{\tau\tau'}_{\Sigma''}(q^2) & = & \frac{q^2}{2m_N^2}c_{10}^\tau c_{10}^{\tau'} + \frac{j_\chi(j_\chi+1)}{12}\left[c_4^\tau c_4^{\tau'} +\frac{q^4}{m_N^4} c_6^\tau c_6^{\tau'}\right],\nn
\eea
where the $c_i^\tau$ are coefficients of the operators $\mathcal{O}_i$ written in the isospin basis $\tau$, rather than the nucleon basis.
The nuclear responses in Eq.~(\ref{eq:Amp}) are defined in terms of the electroweak operators, acting on the nuclear states and summed over multipoles,
\be
W^{\tau\tau'}_{\mc{O}}(q^2) = \sum_{J}^{\infty} \langle j_N \vert \vert \mc{O}_{J;\tau}(q^2)\vert\vert j_N\rangle\langle j_N \vert \vert \mc{O}_{J;\tau'}(q^2)\vert\vert j_N\rangle.
\ee
where, for elastic scattering, the $W_M$ response receives contributions only from the even multipoles, while the $W_{\Sigma'}$ and  $W_{\Sigma''}$ responses only from the odd multipoles. For this analysis the nuclear responses were calculated with the Mathematica package DMFormFactor \cite{Anand:2013yka}. The differential cross section is then,
\bea
\frac{d\sigma}{dE_R} &=& \frac{2 m_T}{(2j_N+1)v^2}  \sum_{\tau=0,1}  \sum_{\tau'=0,1} \left[ R^{\tau\tau'}_{M}(q^2) W^{\tau\tau'}_{M}(q^2) \right.\nn
& & + \left. R^{\tau\tau'}_{\Sigma'}(q^2)  W^{\tau\tau'}_{\Sigma'}(q^2) + R^{\tau\tau'}_{\Sigma''}(q^2)  W^{\tau\tau'}_{\Sigma''}(q^2)\right].
\eea

\subsection{Elastic dark matter-nucleus scattering rates}
First, we will describe our assumptions regarding the local dark matter population encountered in direct detection. A direct detection experiment on earth will encounter a dark matter flux which depends on its local density at the location of the solar system, $\rho_{\chi,\odot}$, and the dark matter velocity distribution, $f(\vec{v}+\vec{v}_e)$, where $\vec{v}_e$ is the Earth's velocity with respect to the galactic rest frame. For our calculations we adopt the value $\rho_{\chi,\odot} = 0.3~{\rm{GeV/cm}}^3$ for the dark matter density, and a Maxwell-Boltzmann velocity distribution with mode at $v_0 = 220$ km/s, given by the circular velocity of the Sun in the galactic rest frame. The velocity distribution is taken to have a cut off at the escape velocity $v_{esc} = 544$ km/s (in the galactic frame):
\be
f(\vec{v}) \propto
\begin{cases}
 \left(e^{-\frac{v^2}{v_0^2}}-e^{-\frac{v_{esc}^2}{v_0^2}}\right) & v \leq v_{esc} \\
0 & v > v_{esc}, \\
\end{cases}
\ee
where the proportionality factor is determined by normalizing the velocity distribution. The elastic WIMP-nucleus differential rate, per unit detector mass, is obtained by averaging over this velocity distribution, 
\be
\frac{dR}{dE_R} = N_T \frac{\rho_{\chi,\odot}}{m_\chi m_T}\int_{v>v_{\mathrm{min}}} \frac{d\sigma}{dE_R} v f(\vec{v}) d^3v,
\ee
where $v_{\min}$ is the minimum speed of an incoming dark matter particle that can produce a given recoil energy, $E_R$, and $N_T$ is the number of target nuclei per unit detector mass. It is customary to place bounds on the dark matter-nucleon cross section at zero momentum transfer, therefore for comparison with other results we calculate the nucleon cross section as,
\be
\sigma_{\chi n} = \frac{{c_i^{n}}^2\mu_{\chi n}^2}{\pi}.
\ee


\section{Dark matter induced Migdal and bremsstrahlung rates}
\label{sec:rates}

In this section, we calculate the rates associated with the Migdal effect and photon bremsstrahlung in the context of dark matter-nucleus scattering. To do so, we must integrate Eq.(\ref{eq:migdal_triple}) and Eq.(\ref{eq:PradlerBremRate}) over the relevant phase-space. In the low momentum-transfer limit relevant for dark matter scattering, these two effects share the same kinematics, which differs from the previous section by the inclusion of an inelastic parameter, $\delta$. For the cases of the Migdal effect and bremsstrahlung, the inelastic parameter is $E_{\rm EM}$ and $\omega$, respectively. The kinematics for both cases and their equivalence is reviewed in Appendix~\ref{app:kinematics}.
The standard relationship between kinematic parameters for dark matter scattering is written in terms of the minimum incoming speed that can produce a given recoil and inelastic energy:
\bea
v_{\rm{min}} = \frac{m_TE_R + \mu_T\delta}{\mu_T\sqrt{2m_TE_R}}.
\eea
As noted in \cite{Dolan:2017xbu}, the endpoints of the phase-space are:
\bea
E_{R,{\rm{max}}} &=& \frac{2\mu_T^2v_{{\rm{max}}}^2}{m_T},\nn
\delta_{\rm{max}} &=& \frac{\mu_Tv_{{\rm{max}}}^2}{2}.
\eea
When the nuclear mass far exceeds the dark matter mass, the global upper bound on the energy deposited is approximately,
\bea
E_{R,\rm{max}} &\approx& 2\left( \frac{m_\chi}{\rm GeV} \right)^2 \left( \frac{\rm GeV}{m_T} \right) \left( \frac{v_{\rm max}^2}{10^{-6}} \right) {\rm keV},\nn
\delta_{\rm{max}} &\approx& \frac{1}{2}\left( \frac{m_\chi}{\rm GeV} \right) \left(\frac{v_{\rm max}^2}{10^{-6}} \right) {\rm keV},\nn
\eea
which allows one to immediately see that $\delta_{\rm{max}} > E_{R,{\rm{max}}}$ due to enhancement by a factor of $m_T / m_\chi > 1$. This general feature is due to the fact that, for a given momentum transfer, more energy can be carried off by light or massless particles. It is this feature that allows the Migdal effect and bremsstrahlung to produce observable signals when the nuclear recoil energy falls below conventional detector thresholds, as the electronic or photonic energy signal can still extend into the observable region. For example, a 1~GeV dark-matter particle incident on xenon with $v_{\rm max}=v_{esc}+v_e$ will give an unmeasurable maximum nuclear recoil of $\sim10^{-2}$~keV, while the maximum inelastic energy is $\sim 3$ keV. Thus, there is a range of dark matter masses where the Migdal effect can provide new discovery channels or a means of constraining dark matter interactions via non-observation.

\subsection{Detected energy}

As the nuclear recoil energy is well below a keV for the light dark matter masses (below a few GeV, dependent on target mass) we will be considering, it will be undetectable. The physical quantity we want to calculate is the differential rate with respect to the detected energy, $E_{\rm det}$. We thus integrate Eq.~(\ref{eq:migdal_triple}) and Eq.~(\ref{eq:PradlerBremRate}) over all possible nuclear recoil energies and average over the incoming dark matter speeds. These integrals can be performed in either order, so long as one takes the appropriate phase space limits.

For the elastic nuclear recoil events the detected electron-equivalent energy is (i.e. the quenched nuclear recoil energy): $E_{\rm det}= Q(E_R) E_R$, where $Q(E_R)$ is the recoil energy dependent quenching factor. Using this relationship allows us to compare the nuclear and electronic recoil spectra as seen by a detector. The quenching factor is target dependent and is usually in the range of 0.1-0.3. For the case of xenon, various measurements of the quenching factor have been made with a range of outcomes (see Fig.~1 of \cite{Aprile:2011hi}). For xenon, which will be used for to set bounds in this analysis, we use an energy dependent form of the quenching factor given by Lindhard theory~\cite{osti_4153115}. For the other targets, where we only provide rate calculations, for simplicity we use constant quenching factors of 0.25, 0.15 and 0.10 for argon, germanium and sodium targets, respectively~\cite{Gastler:2010sc,Scholz:2016qos,Stiegler:2017kjw}. We note that this scaling is approximate and for comparative purposes only. For both the Migdal and bremsstrahlung events the detected energy is from electronic recoil and is therefore not quenched, given by $E_{\rm det}=E_{\rm n\ell}+E_e$ and $E_{\rm det} = \omega$, respectively.

Using a liquid xenon detector as an example of how these effects manifest in an experimental setting, a photon (produced directly from bremsstrahlung or through atomic de-excitation) with energy in the range of hundreds of eV, up to a few keV, will not propagate far in liquid xenon and can be assumed to convert all of its energy into ionization of valence electrons. Such an event will produce a large signal appearing in the electronic recoil band, and be detected with high efficiency. That being the case, any such signal will have to contend with the higher rate of electronic recoil backgrounds. 

We will now show that the Migdal rate dominates over the bremsstrahlung rate for a variety of targets and interactions. This could plausibly be anticipated (and was remarked upon in \cite{Ibe:2017yqa}) via an examination of Eq.~(\ref{eq:migdal_triple}) and Eq.~(\ref{eq:PradlerBremRate}), where we see that the bremsstrahlung rate features an additional suppression factor of $E_R/m_T$ in comparison with the Migdal rate.


\subsection{Rates for different targets}

We calculate the rates of elastic nuclear recoil, the Migdal effect, and bremsstrahlung induced in a representative sample of targets used in current and future dark matter detection which includes liquid xenon (LXe), liquid argon (LAr), sodium crystals, and germanium crystals. We perform the calculations for our four chosen operators $\mathcal{O}_1$, $\mathcal{O}_{4}$, $\mathcal{O}_{6}$ and $\mathcal{O}_{10}$. The momentum-independent operators $\mathcal{O}_{1}$ and $\mathcal{O}_{4}$ are the standard spin-independent and spin-dependent operators, typically dubbed SI and SD, respectively. The $\mathcal{O}_{6}$ and $\mathcal{O}_{10}$ operators are chosen as representatives of operators with momentum dependence of $\mathcal{O}(q^2/m_T^2)$ and $\mathcal{O}(q/m_T)$, respectively (grouping of operators in this fashion is standard practice, see \cite{Catena:2015vpa,Gluscevic:2015sqa,Dent:2016iht,Dent:2016wor} for example). The rate of spin-dependent scattering is highly dependent on the isotopic makeup of the target, since the addition or subtraction of single neutrons can open or close nuclear shells. Therefore, we take the weighted average over the natural abundance of the isotopes in the target for all of our calculations. In this analysis we have neglected atomic effects for bremsstrahlung. Including these effects would modestly reduce the rates by a common amount for all cases considered and serves to provide a cut off at low energy (see Ref.~\cite{Kouvaris:2016afs} for a detailed treatment).

\begin{figure*}[htb]
\begin{tabular}{cc}
\includegraphics[width=6cm]{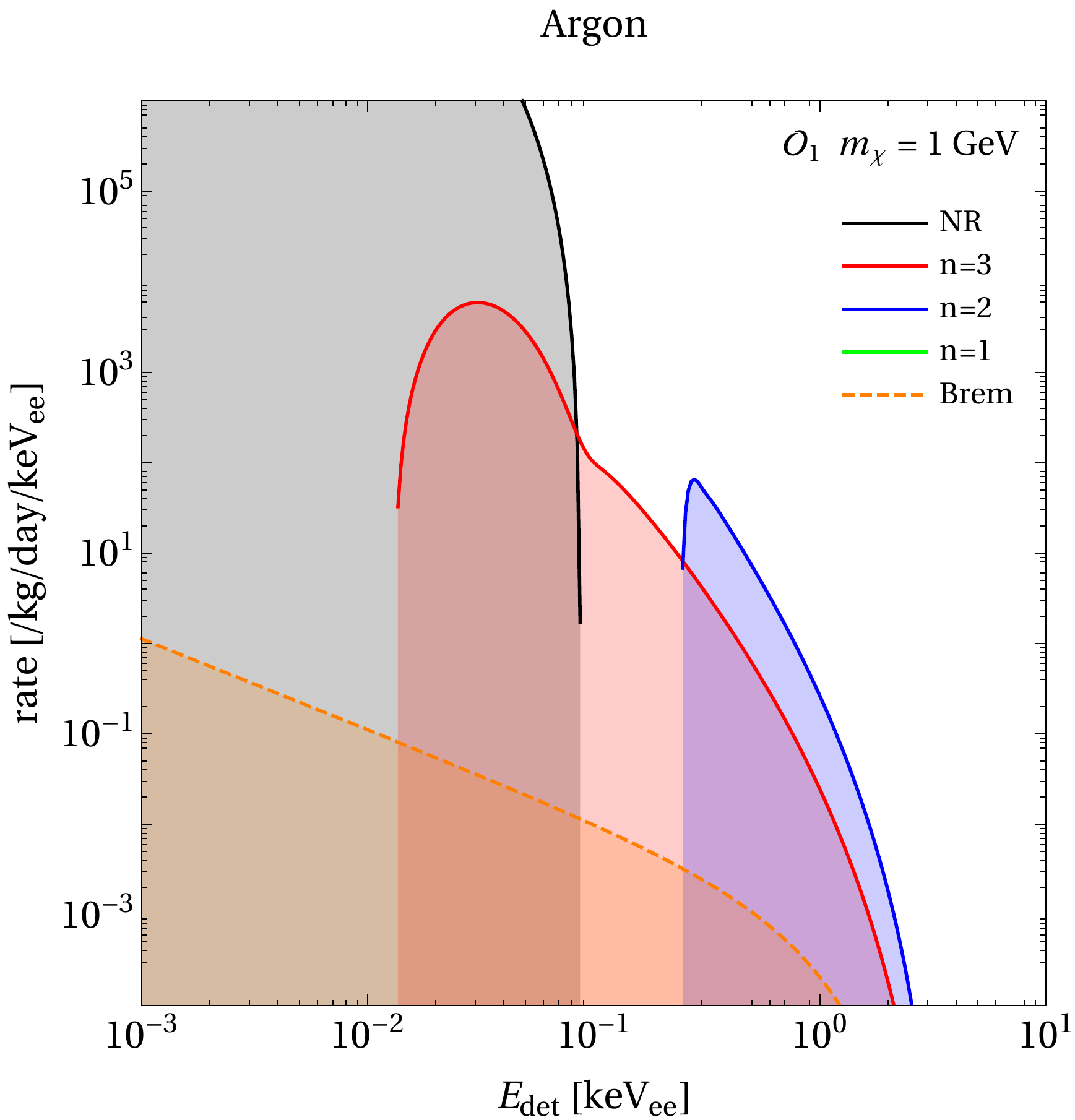} & \includegraphics[width=6cm]{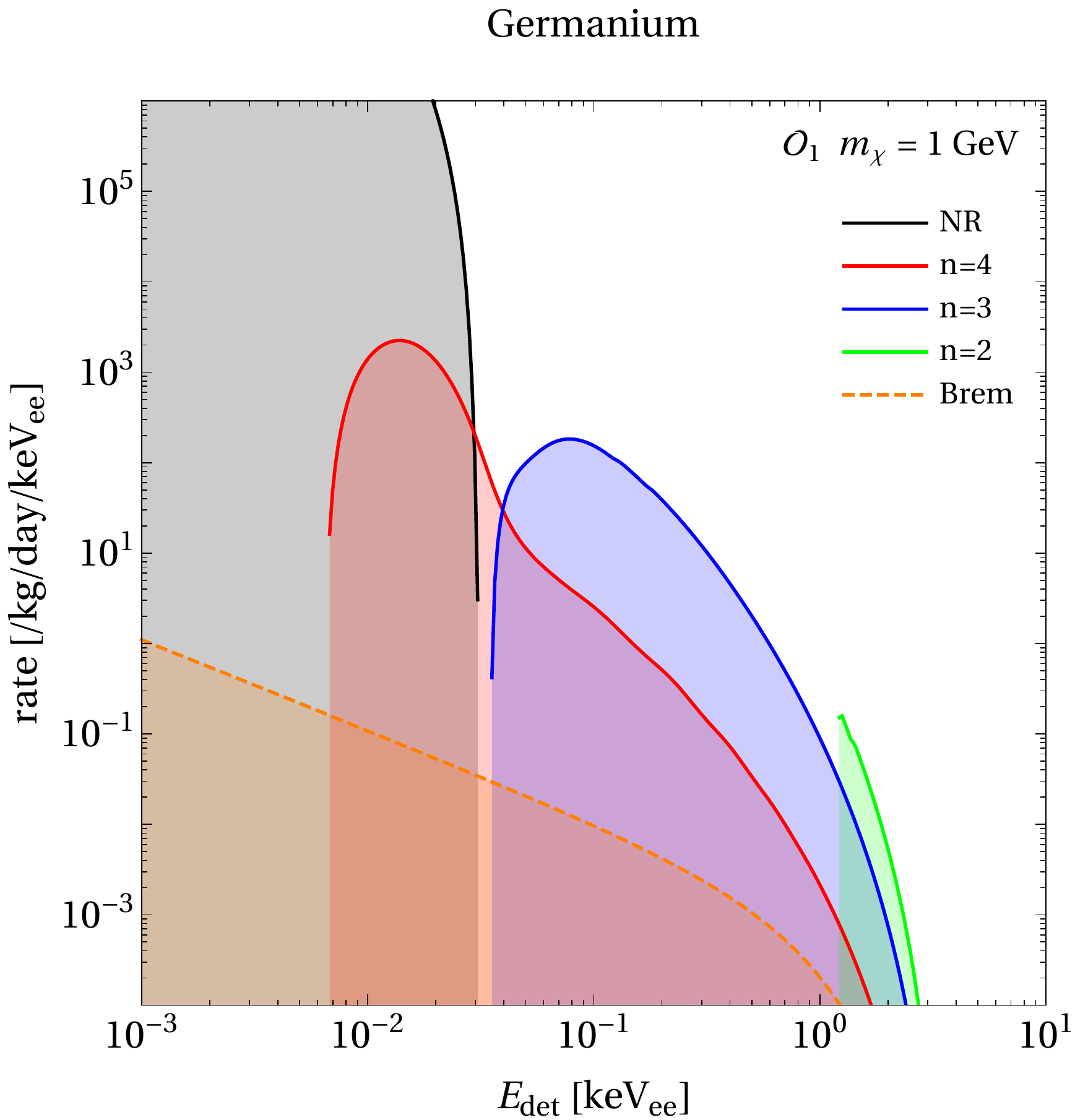}
\\
\includegraphics[width=6cm]{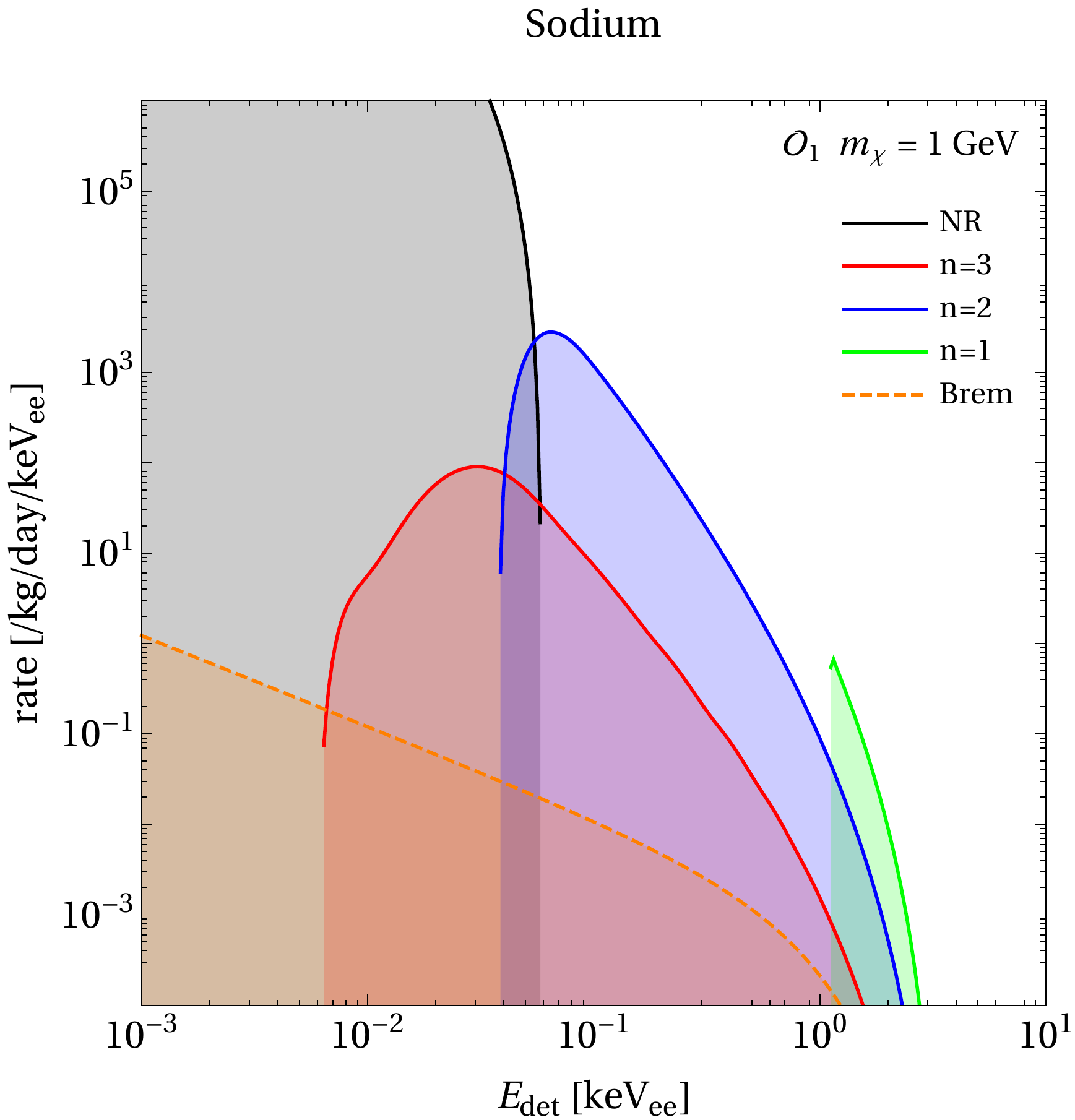} & \includegraphics[width=6cm]{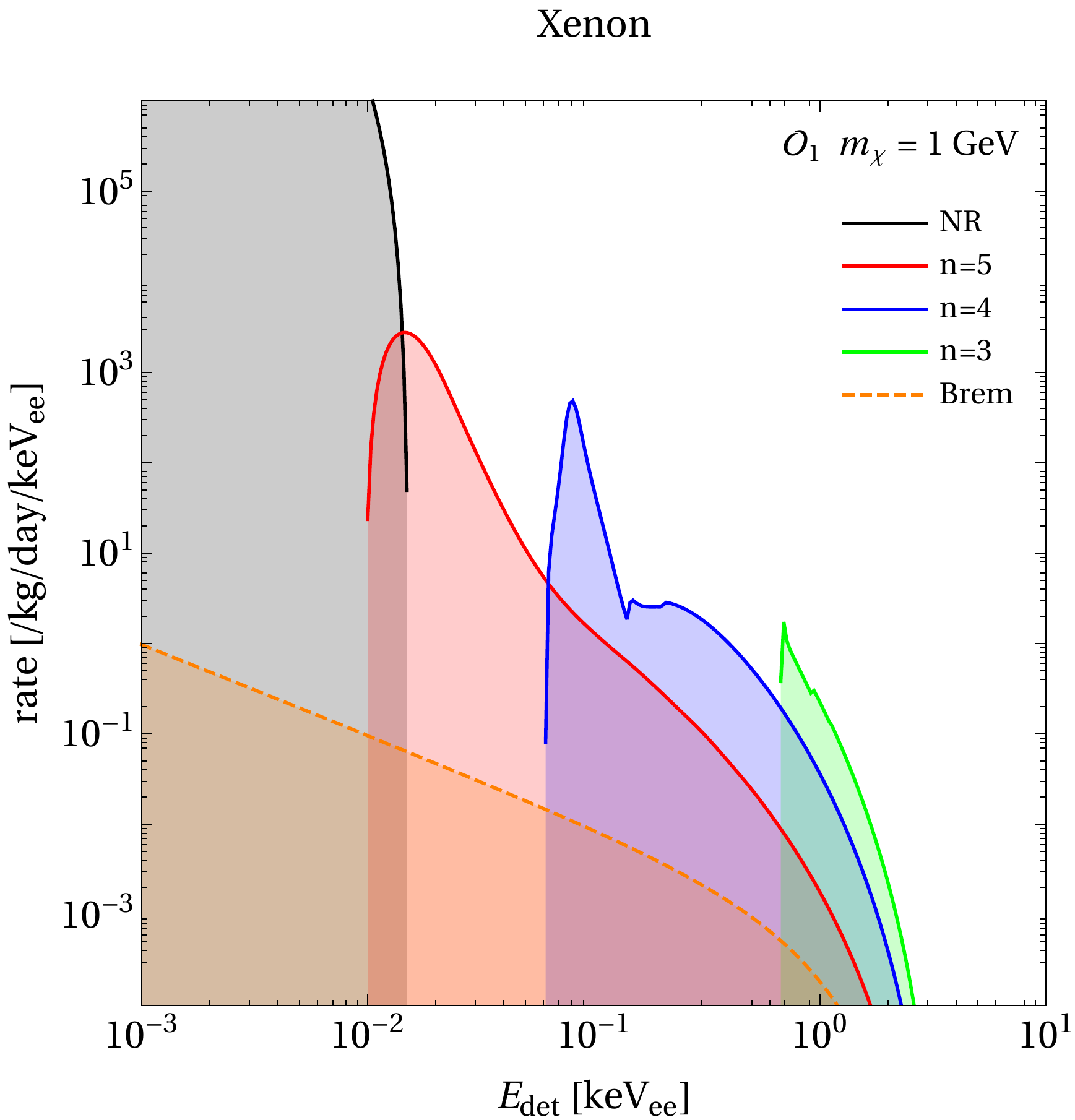}
\end{tabular}
\caption{The rate for the Migdal effect (depicted by the colored lines associated with various atomic energy levels denoted by $n$) and bremsstrahlung (dashed orange line) induced by a 1 GeV mass dark matter particle interacting with the nucleus through the spin- and momentum-independent operator $\mathcal{O}_1 = \mathbb{1}_\chi\mathbb{1}_N$ for the target elements argon (top left), germanium (top right), sodium (bottom left), and xenon (bottom right). The rate for the standard nuclear recoil is depicted by the black line.}
\label{fig:O1_1}
\end{figure*}

In Fig.~\ref{fig:O1_1}, we show the rates as a function of energy (in units of keV electron-equivalent, keV$_{\rm ee}$) induced by the $\mathcal{O}_1$ interaction operator for argon (top left), germanium (top right), sodium (bottom left), and xenon (bottom right) for an incident dark matter particle with a 1 GeV mass (the rates for all four interaction operators with $m_\chi = 2$ GeV and 0.5 GeV and using LXe as the target are displayed in Appendix~\ref{app:2and5rates}). For comparison, in each case the interaction strength is scaled such that $10^6$ nuclear recoils are expected. The solid black line represents the spectrum for nuclear recoil, the solid color lines give the Migdal effect rates for the various electron energy levels designated by $n$, and the dashed orange line represents the spectrum for the photon emitted via bremsstrahlung.  One notes several features common to all targets. First, the rates for both the Migdal effect and the bremsstrahlung process are strongly suppressed compared to the peak rate for nuclear recoils, with the bremsstrahlung rate being dominated by the Migdal effect across all energies and targets. Secondly, the Migdal rate naturally differs for each atomic target due to the different electron energy levels that are occupied. Finally, an essential point to note is: though the Migdal rate is significantly lower than the peak nuclear recoil rate, it is the dominant rate for a 1 GeV dark matter particle above 100 $\rm{keV}_{\rm{ee}}$ for all targets. For many current detector technologies, their detection thresholds lie squarely in this region, leading to the possibility of utilizing the Migdal effect to lower the reach of detectors into the $\lesssim1 {\rm{GeV}}$ dark matter mass window.

\begin{figure*}[htb!]
\begin{tabular}{ccc}
\includegraphics[width=5.8cm]{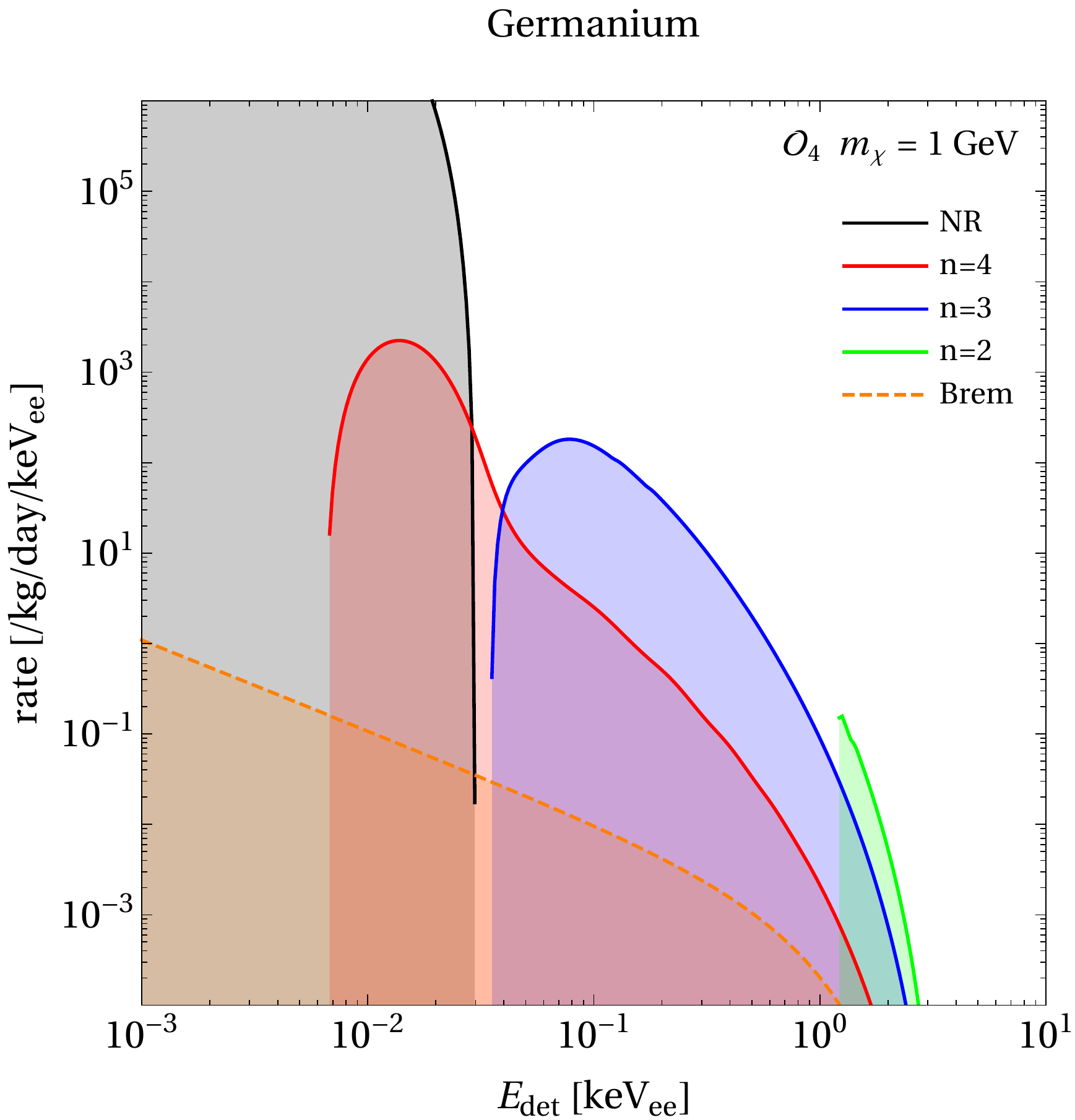}
&
\includegraphics[width=5.8cm]{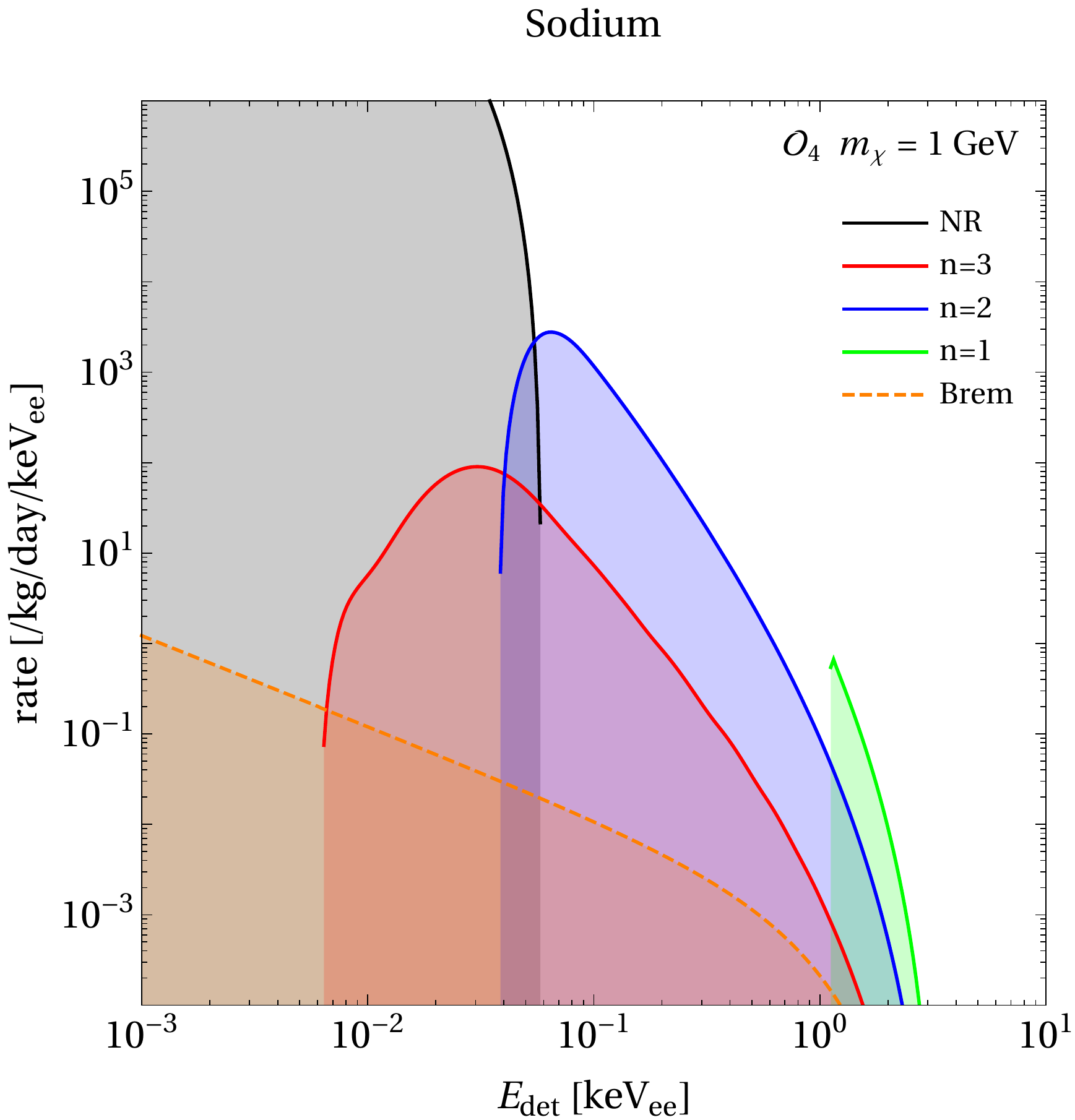} 
&
 \includegraphics[width=5.8cm]{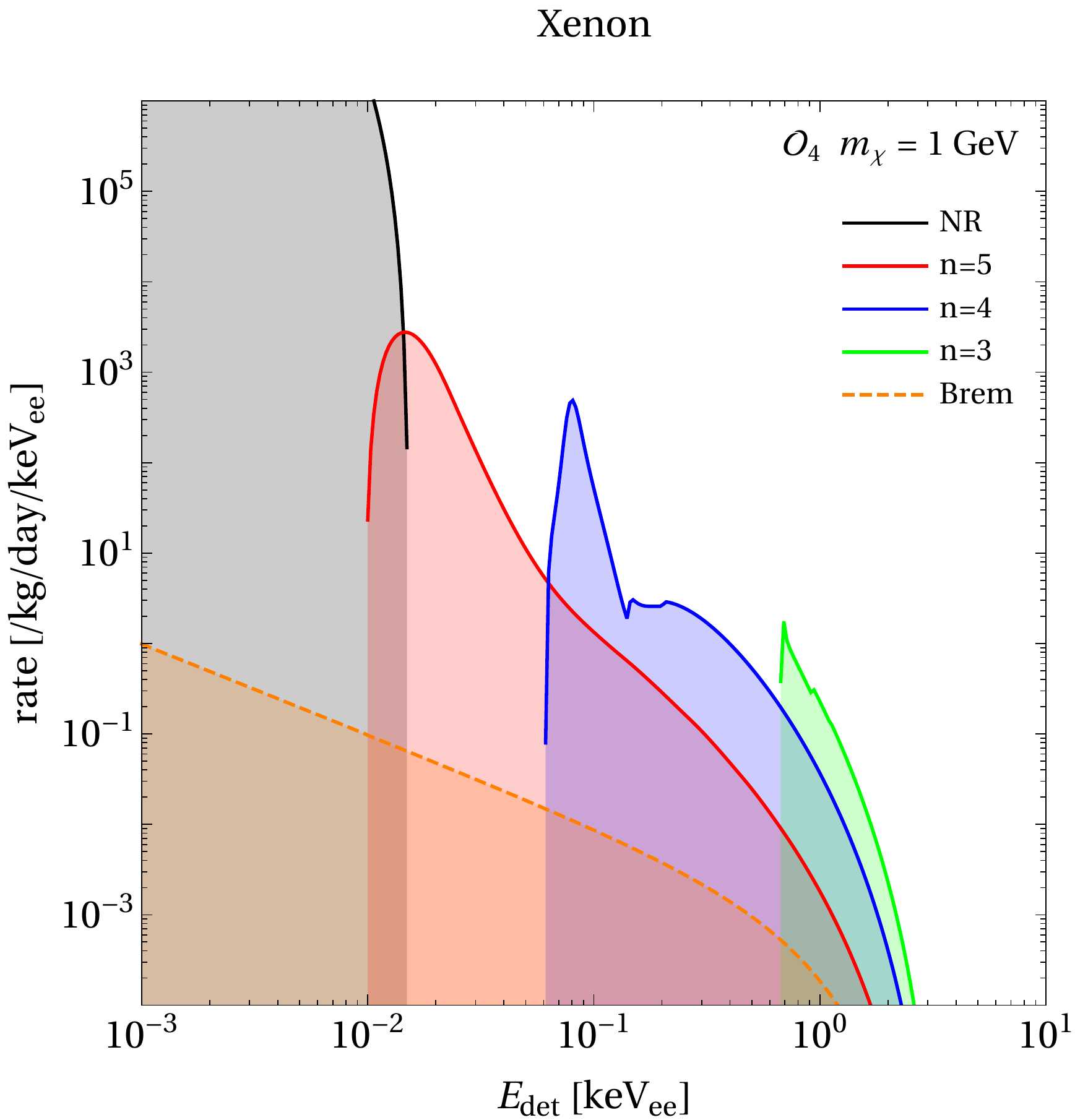}\\
 \includegraphics[width=5.8cm]{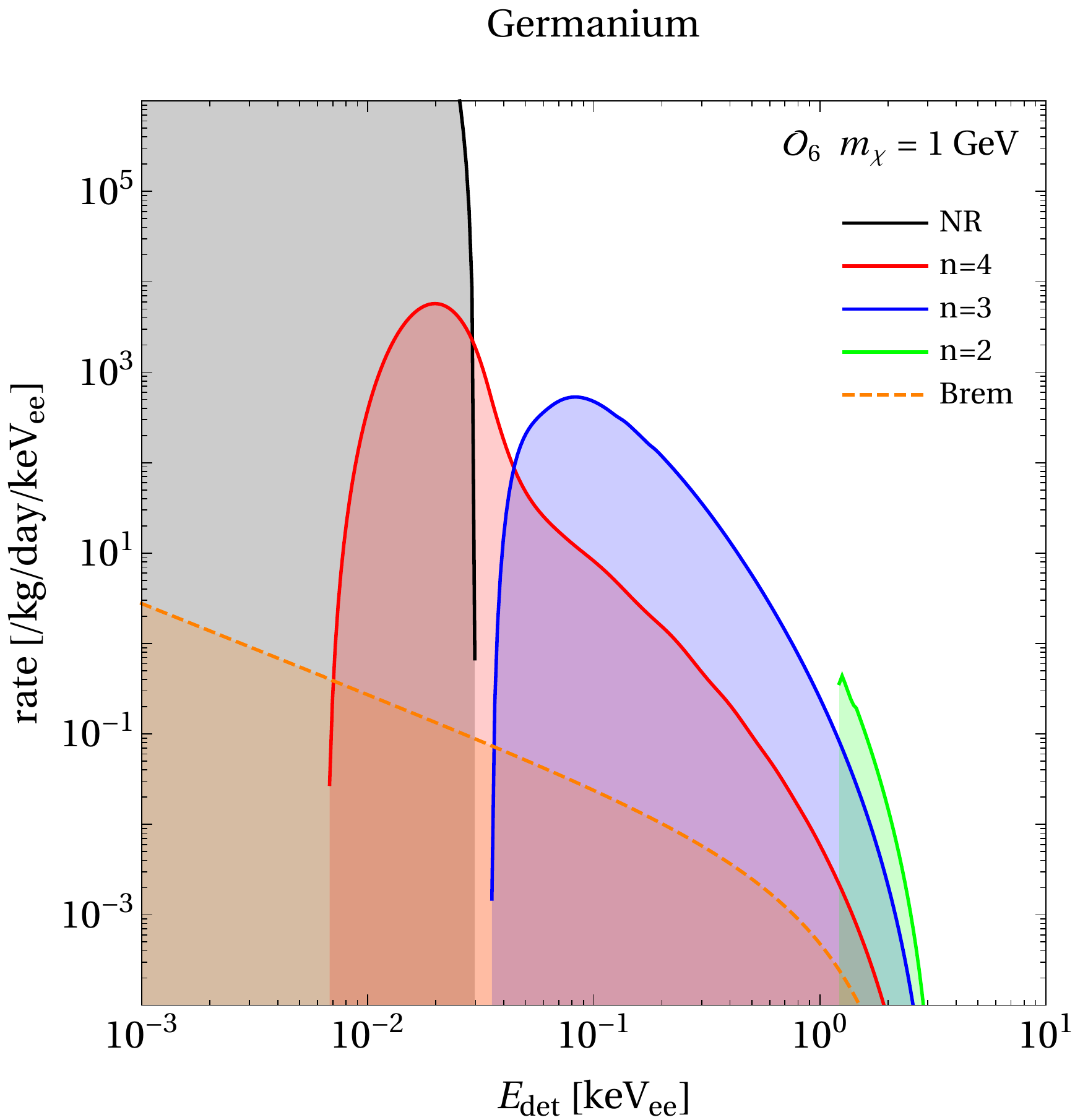}
&
\includegraphics[width=5.8cm]{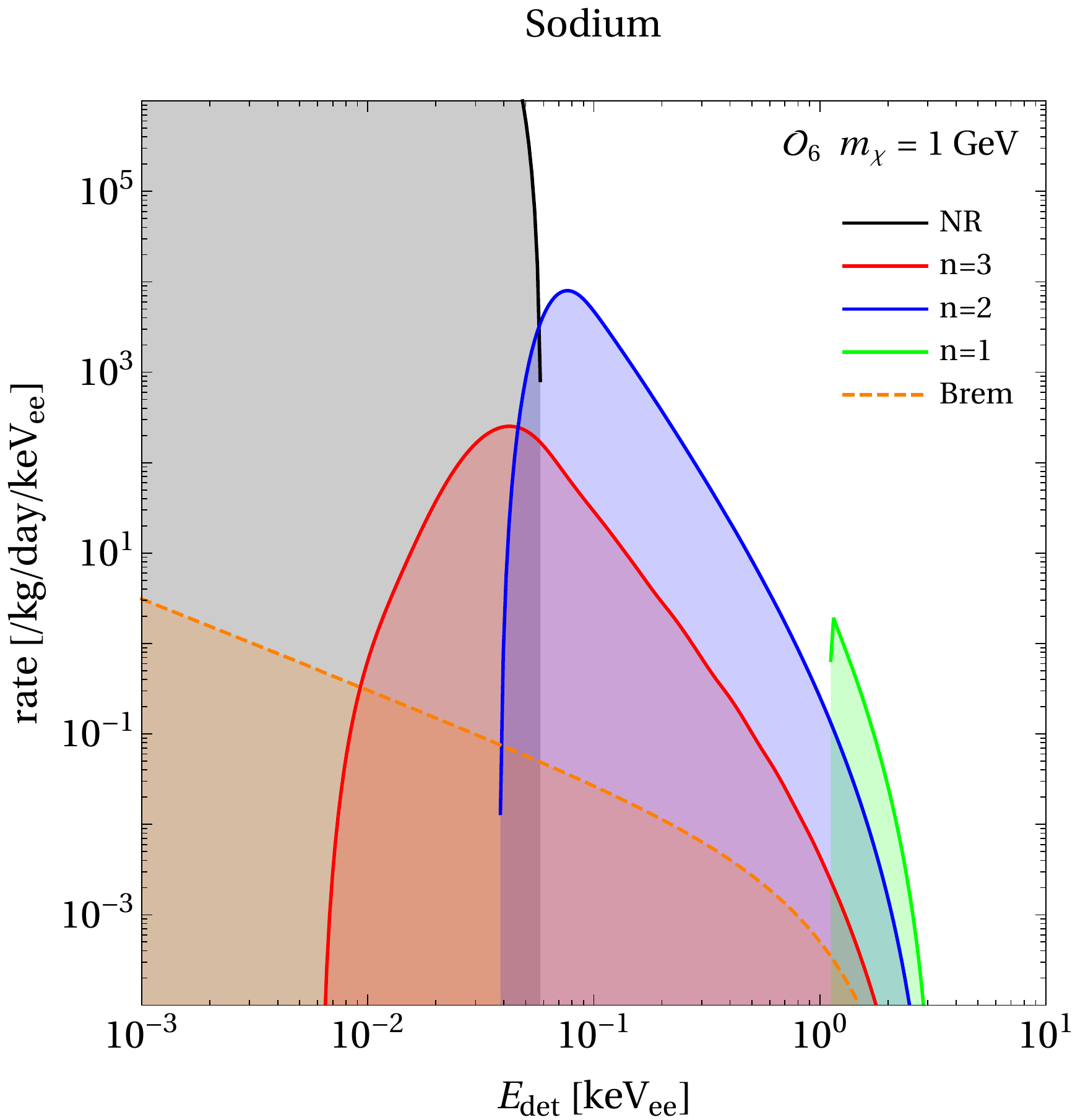} 
&
 \includegraphics[width=5.8cm]{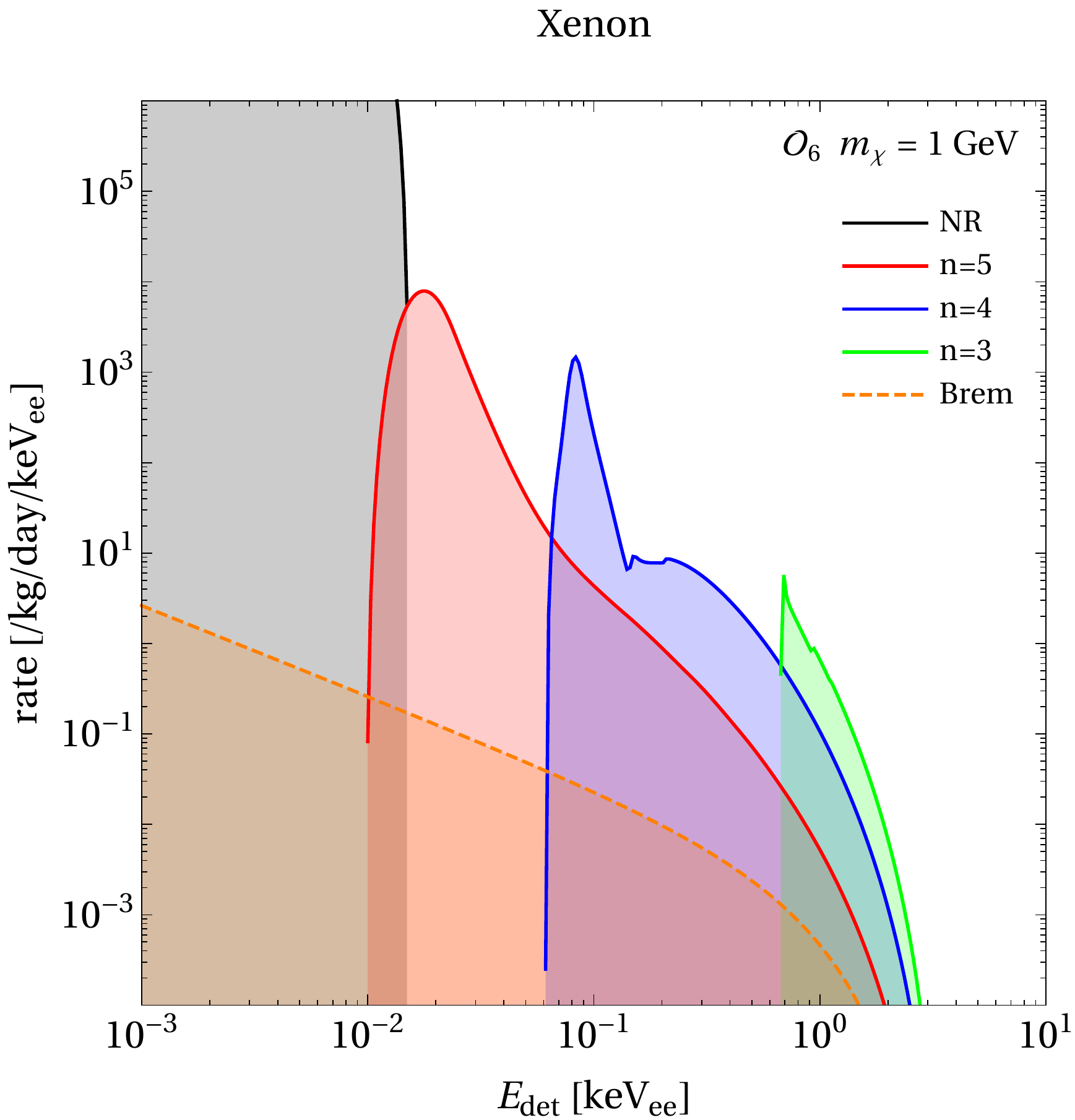}\\
 \includegraphics[width=5.8cm]{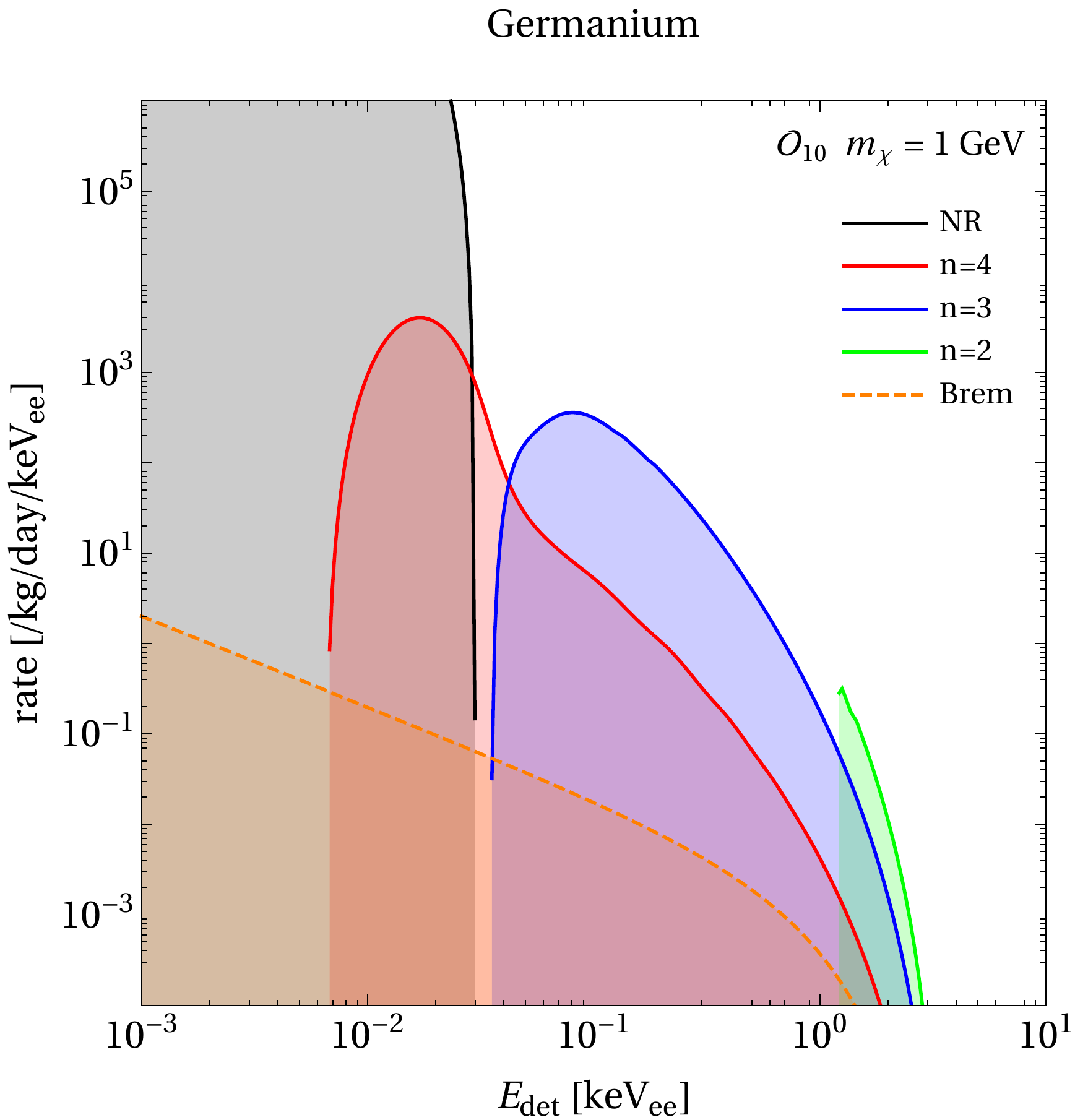}
&
\includegraphics[width=5.8cm]{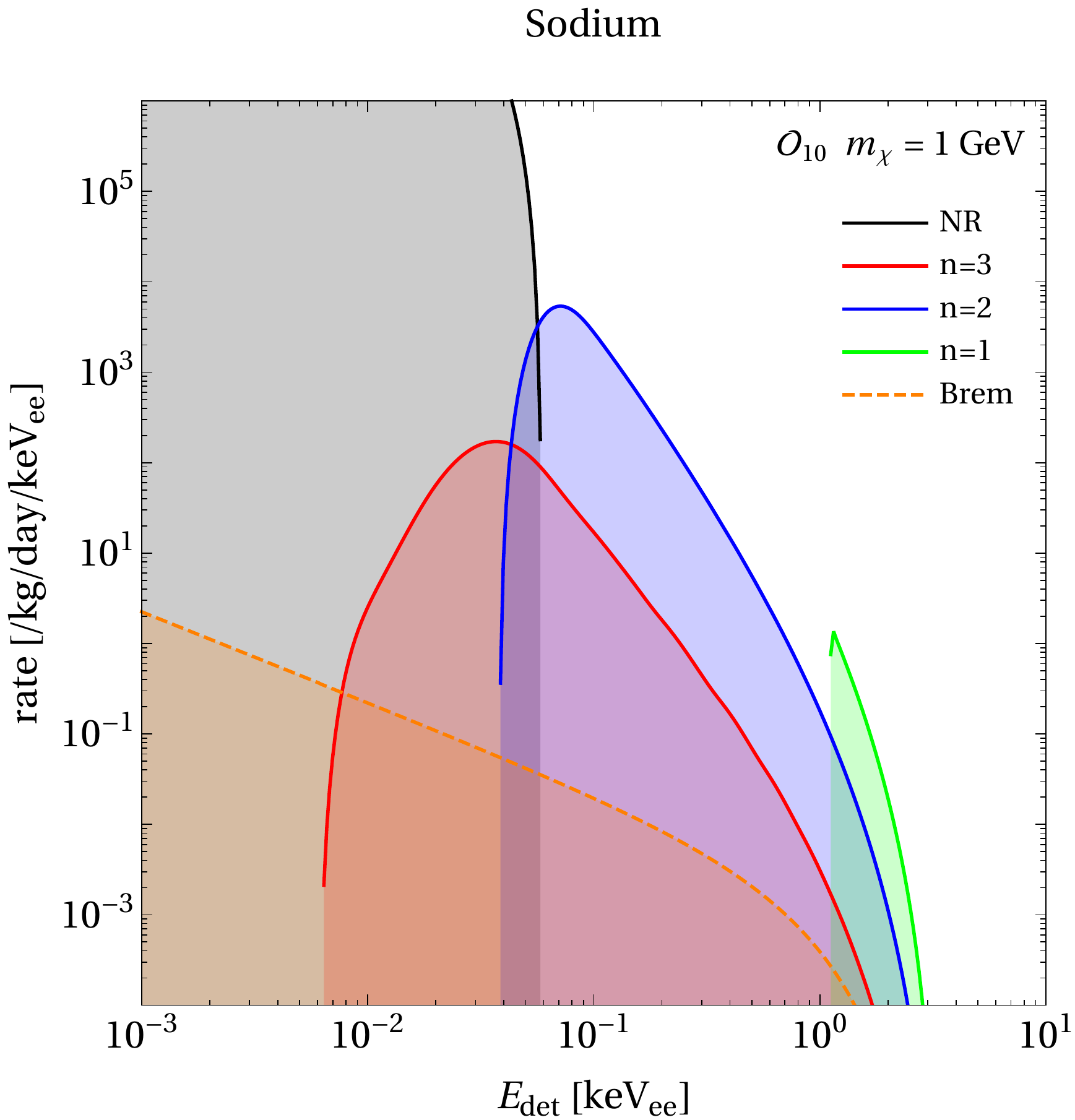} 
&
 \includegraphics[width=5.8cm]{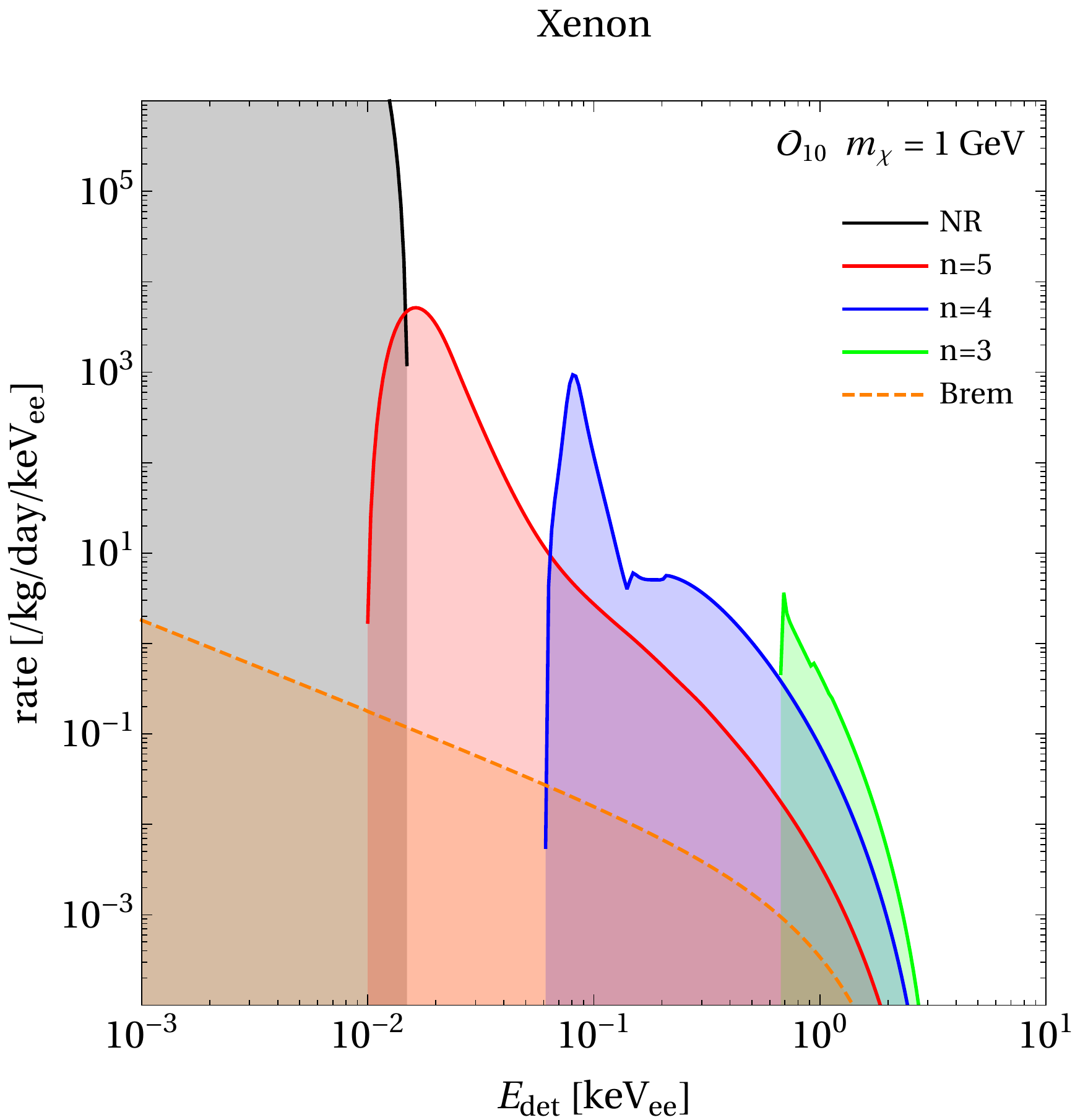}
\end{tabular}
\caption{The rate for the Migdal effect (depicted by the colored lines associated with various atomic energy levels denoted by $n$) and bremsstrahlung (dashed orange line) induced by a 1 GeV mass dark matter particle interacting with the target elements germanium (left), sodium (center), and xenon (right), for the spin-dependent but momentum-independent operator $\mathcal{O}_4 = \vec{S}_\chi \cdot \vec{S}_N$ (top row), the spin- and momentum-dependent operator $\mathcal{O}_6 = \left(\frac{\vec{q}}{m_T}\cdot\vec{S}_{\chi}\right)\left(\frac{\vec{q}}{m_T}\cdot\vec{S}_{N}\right)$ (middle row) and the spin and momentum dependent operator $\mathcal{O}_{10} = \mathbb{1}_\chi \left(i\frac{\vec{q}}{m_T}\cdot\vec{S}_N\right)$ (bottom row).
The rate for the standard nuclear recoil is depicted by the black line.}
\label{fig:O4O6O10_1}
\end{figure*}

We see this type of scenario play out for each of the operators studied, as depicted in Figs.~\ref{fig:O1_1},~\ref{fig:O4O6O10_1}, and~\ref{fig:otherMasses} (in Appendix \ref{app:2and5rates}) . In every combination of operator and target studied, the bremsstrahlung rate is subdominant to the Migdal effect, which is the leading signal for a significant portion of the $\lesssim$ keV recoil range. 
Since the bremsstrahlung rates are sub-dominant, we adopt the reasonable simplification of neglecting them for any detailed experimental calculations.
For the spin-dependent operators, $\mathcal{O}_4$, $\mathcal{O}_6$, and $\mathcal{O}_{10}$, only germanium, sodium, and xenon responses are shown. This is because the argon target studied is ${}^{40}$Ar, which is insensitive to spin-dependent interactions. For each of the operators, the Migdal rate is almost indistinguishable. This is because the Migdal rate at a given $E_{\mathrm{det}}$ is integrated over all $E_{R}$, and so only a small residual dependence on the operator remains. This is most visible at small $E_{\mathrm{det}}$, where the bounds of the $E_{R}$ integral are restricted to low recoils, where the NR rate differs the most between operators.


\section{Bounds on EFT operators from the Migdal effect}
\label{sec:limits}

Now that we have calculated the rates and determined that the bremsstrahlung process can be neglected relative to the Migdal effect in all of the cases under inspection, we will use the Migdal effect results to place new limits on a variety of EFT operators for low mass dark matter. Our bounds will be placed using XENON1T, a liquid xenon detector of the dual-phase (liquid and gas) time projection chamber (TPC) type. In these detectors nuclear and electronic recoils produce both prompt scintillation and ionization. The former signal (S1) is measured directly, and the latter is measured as a proportional signal (S2) when the drifted electrons are extracted into the gas phase. Electronic recoils produce events with a much larger S2 to S1 ratio when compared to nuclear recoils. This fact is the origin of nuclear/electronic recoil discrimination in xenon TPC's (argon TPC's can additionally make use of pulse shape discrimination~\cite{Agnes:2017grb}).

The XENON collaboration have performed several analyses of their 1.3 tonne years of exposure, providing world-leading bounds on dark matter interactions. The standard analysis makes use of both S1 and S2 signals to perform a low background search (enabled by discrimination) for NR due to WIMPs~\cite{Aprile:2018dbl}. Alternately, one can analyse the same data using the S2 signal alone. This `S2-only' analysis sacrifices background discrimination and live-time in exchange for a lower threshold, this enables stronger bounds to be set on low-mass WIMPs~\cite{Aprile:2019xxb}. 

In this work we calculate bounds on each of the operators via to the Migdal effect using the S2-only analysis, since this provides the best sensitivity available. Additionally we calculate bounds on the operators using the S1S2 and S2-only analyses directly from the nuclear recoil channel. The calculation was carried out using a cut and count method. This method does not require a thorough detector simulation, greatly simplifying the computation while still capturing the general detector characteristics. A more advanced analysis could leverage the location of the Migdal events in the S1/S2 space; in \cite{McCabe:2017rln}, however, this was found to only have a small effect. The energy dependent efficiency of nuclear and electronic recoils detection was applied for both analyses (see Fig. 1 of \cite{Aprile:2019xxb}). The 90\%CL was found using the profile likelihood ratio~\cite{Cowan:2010js}, based on Poisson likelihood. For the S1S2 analysis we use a single bin where $n_{\mr{exp}}=7.36$ and $n_{\mr{obs}}=14$). For the S2-only analysis a large number of events were observed at low energies, to improved the bounds we break the data into two bins (around the point $E_R = 0.3 \mr{keV_{ee}}$) with $n_{\mr{exp}}=\{38.8,26.5\}$ and $n_{\mr{obs}}=\{153,39\}$. 

\begin{figure}[htb!]
 \includegraphics[width=8cm]{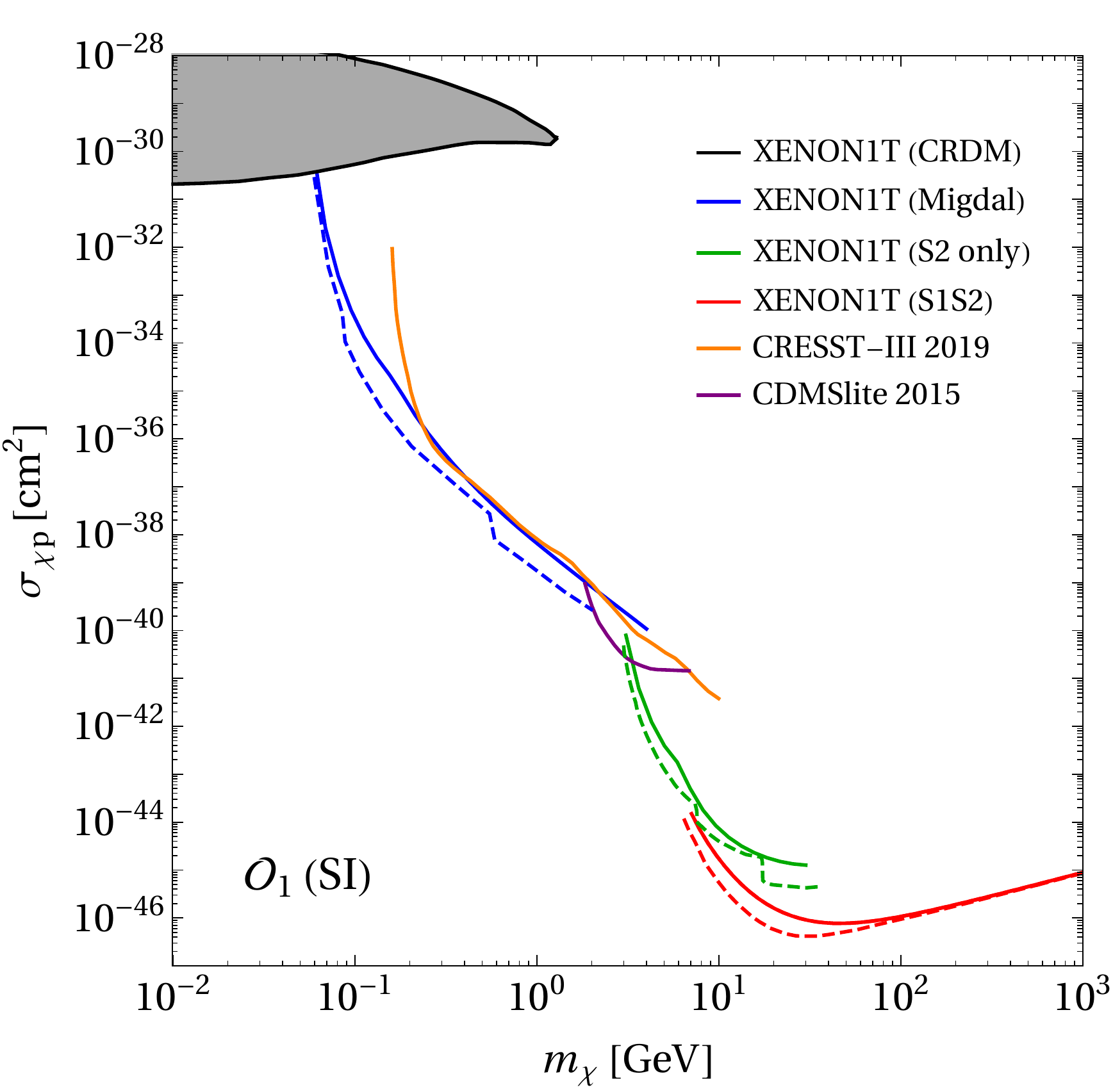}
\caption{The bounds from XENON1T on the spin-independent WIMP-nucleon cross section from the S1S2 NR analysis (red), S2-only NR analysis (green), the S2-only Migdal analysis (blue), and `cosmic ray dark matter' (black)~\cite{Bringmann:2018cvk}. The dashed curves show official limits from the XENON1T collaboration. Bounds from CDMSlite~\cite{Agnese:2017jvy} and CRESST-III~\cite{Abdelhameed:2019hmk} are also shown for comparison.}
\label{fig:siExclusion}
\end{figure}

The bounds on the spin-independent cross section are shown in Fig.~\ref{fig:siExclusion}. For the spin-independent case, the neutron and proton cross sections are assumed to be equal (no isospin violation). The broad agreement between our nuclear recoil limits and the official limits from XENON1T verifies that our approach is accurately modelling the experiment. Though we do not calculate the effect here it should be noted that, for large enough cross-sections, underground detectors will lose sensitivity to scattering signals from incoming dark matter as the incident flux is attenuated due to scattering in the earth \cite{Kavanagh:2016pyr,Emken:2017qmp,Emken:2018run,Emken:2019hgy}.

\begin{figure*}[htb]
\begin{tabular}{ccc}
\includegraphics[width=5.8cm]{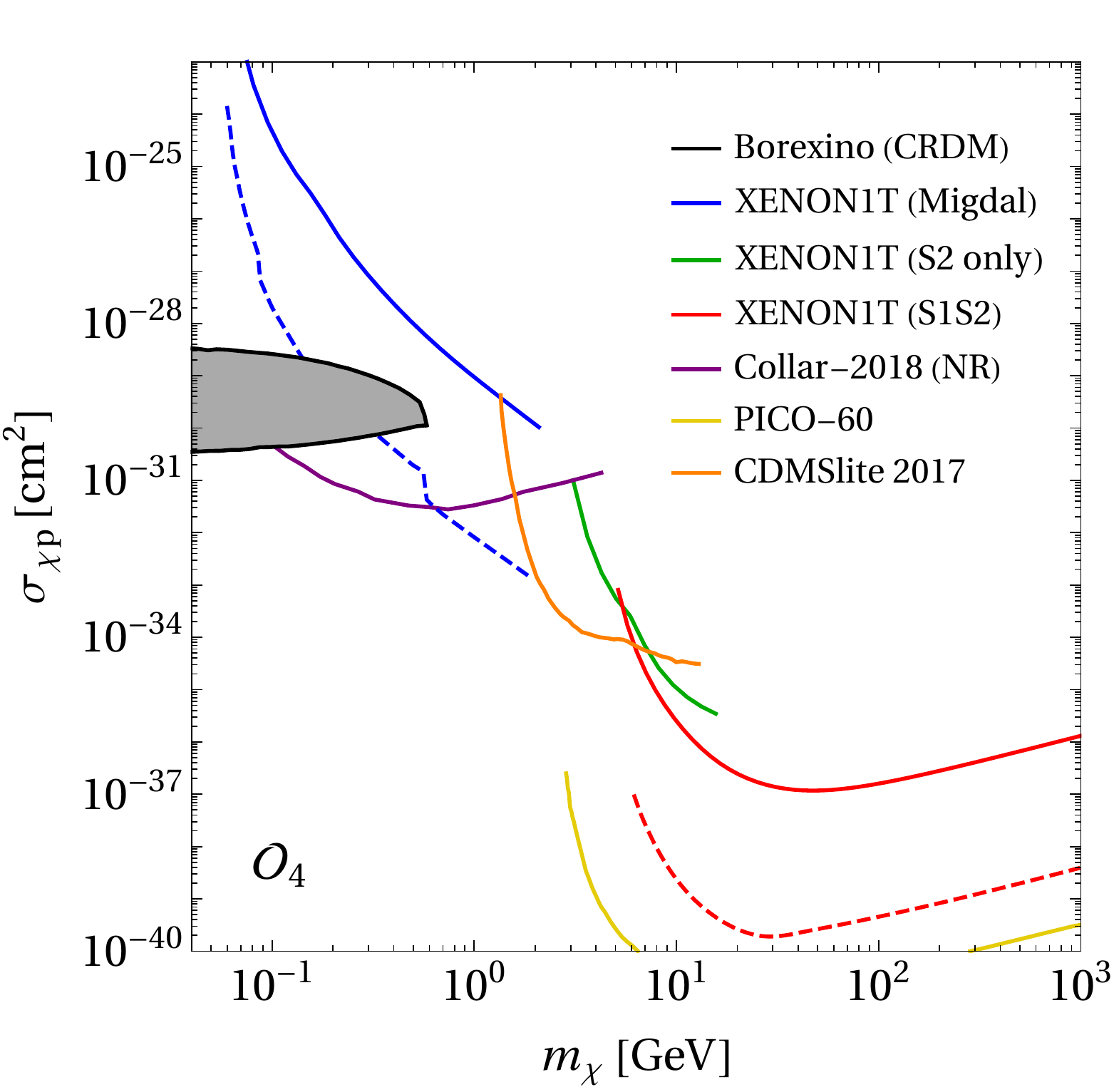}
&
\includegraphics[width=5.8cm]{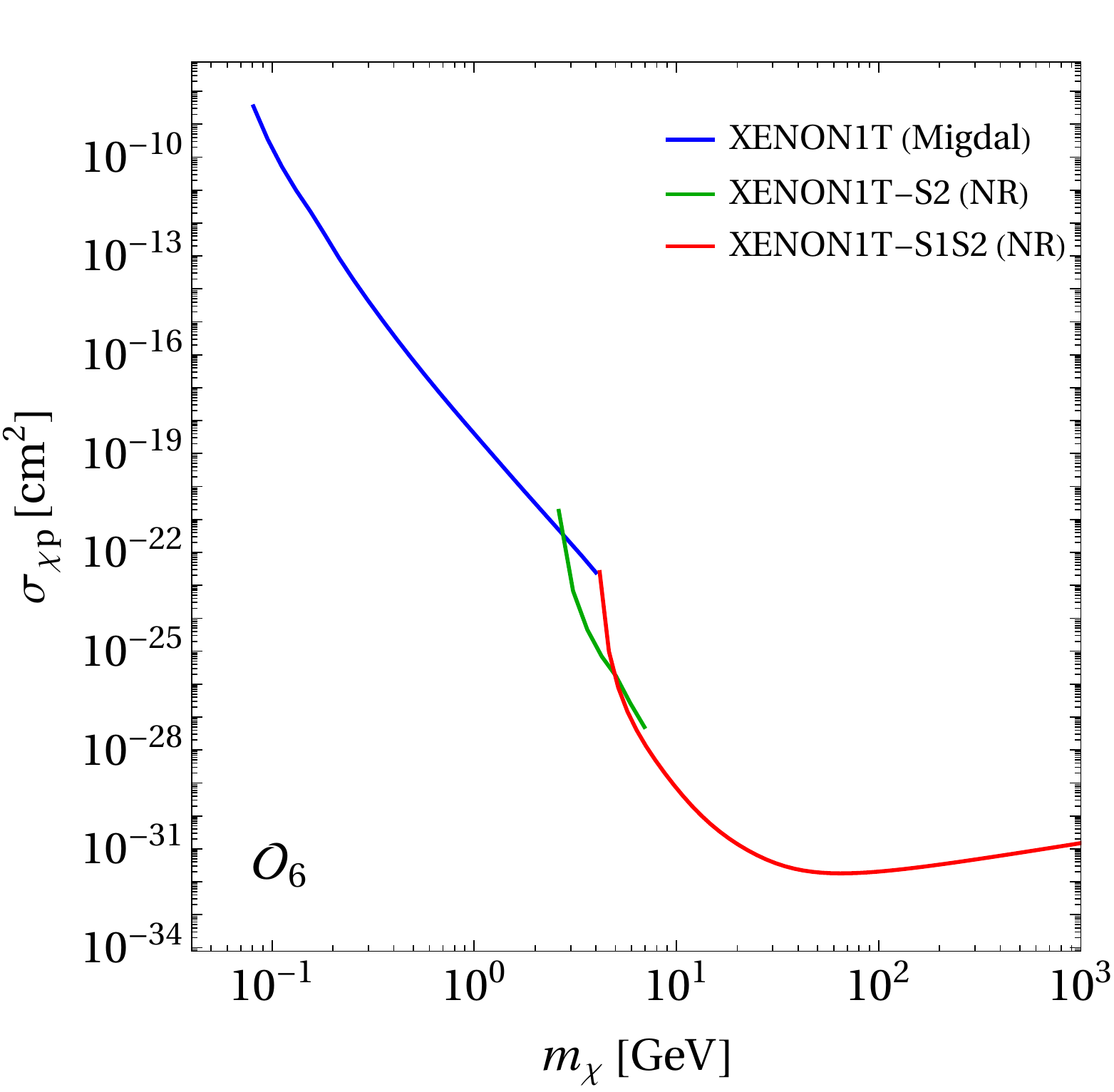} 
&
 \includegraphics[width=5.8cm]{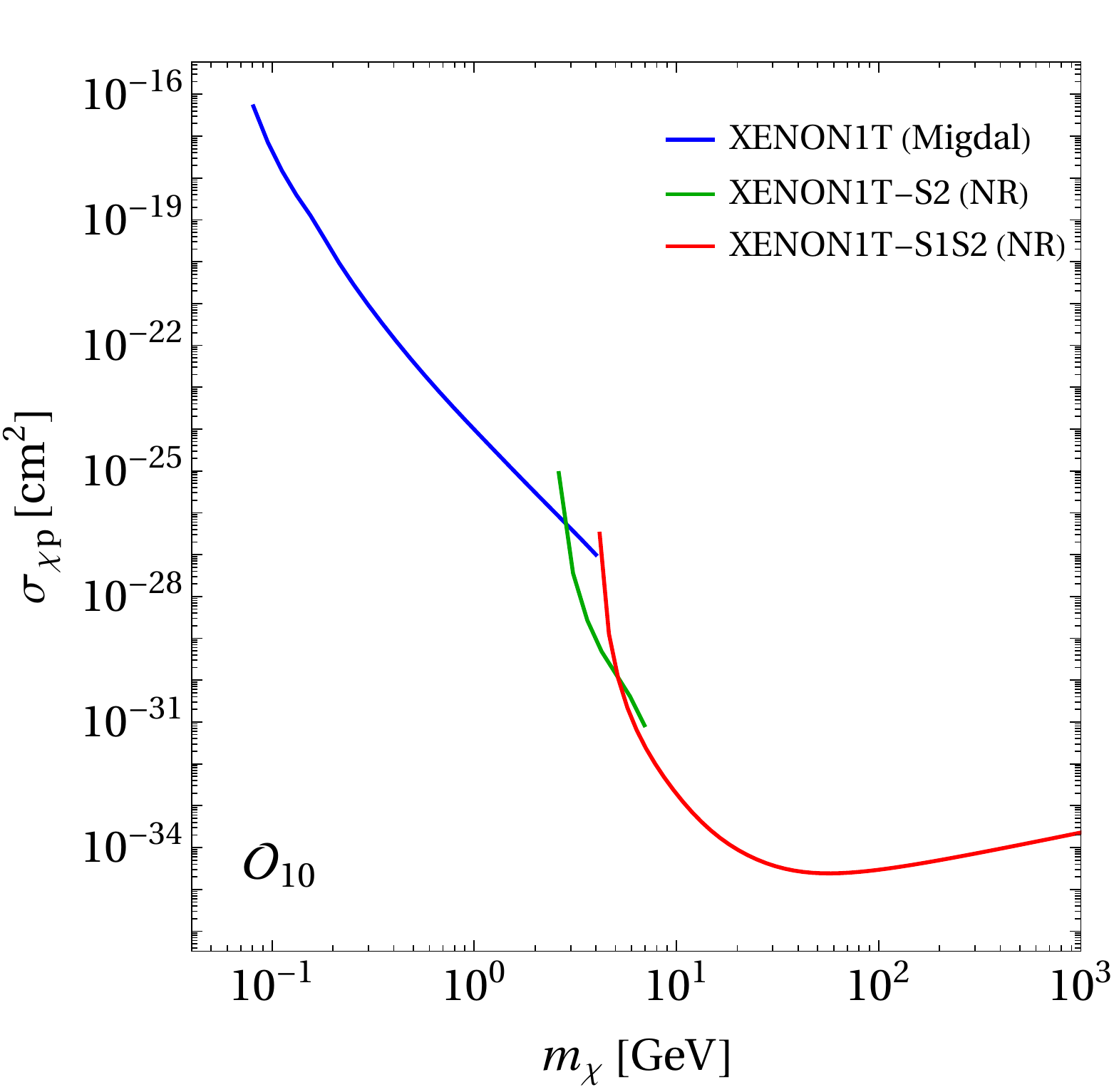}\\
\includegraphics[width=5.8cm]{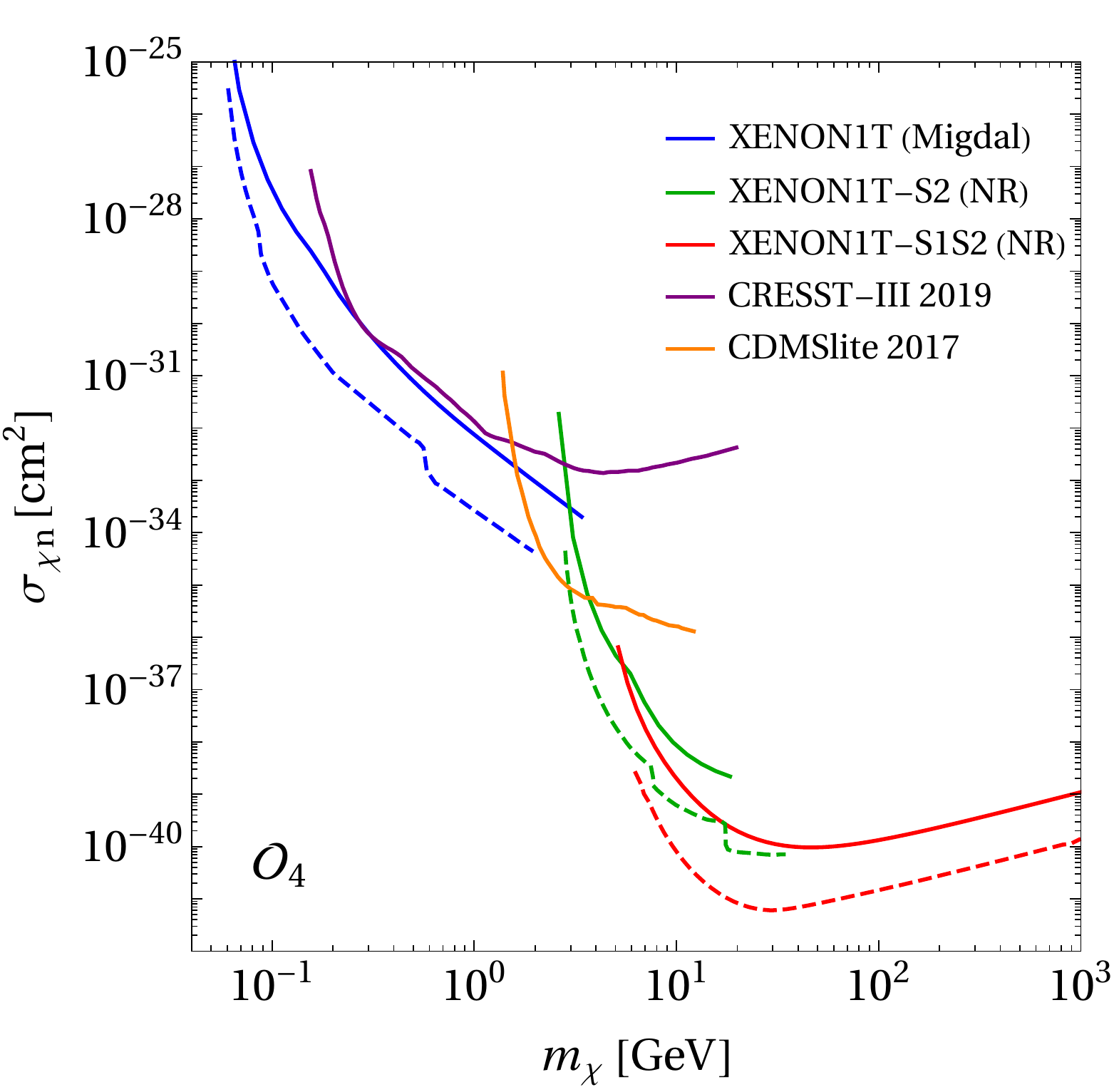}
&
\includegraphics[width=5.8cm]{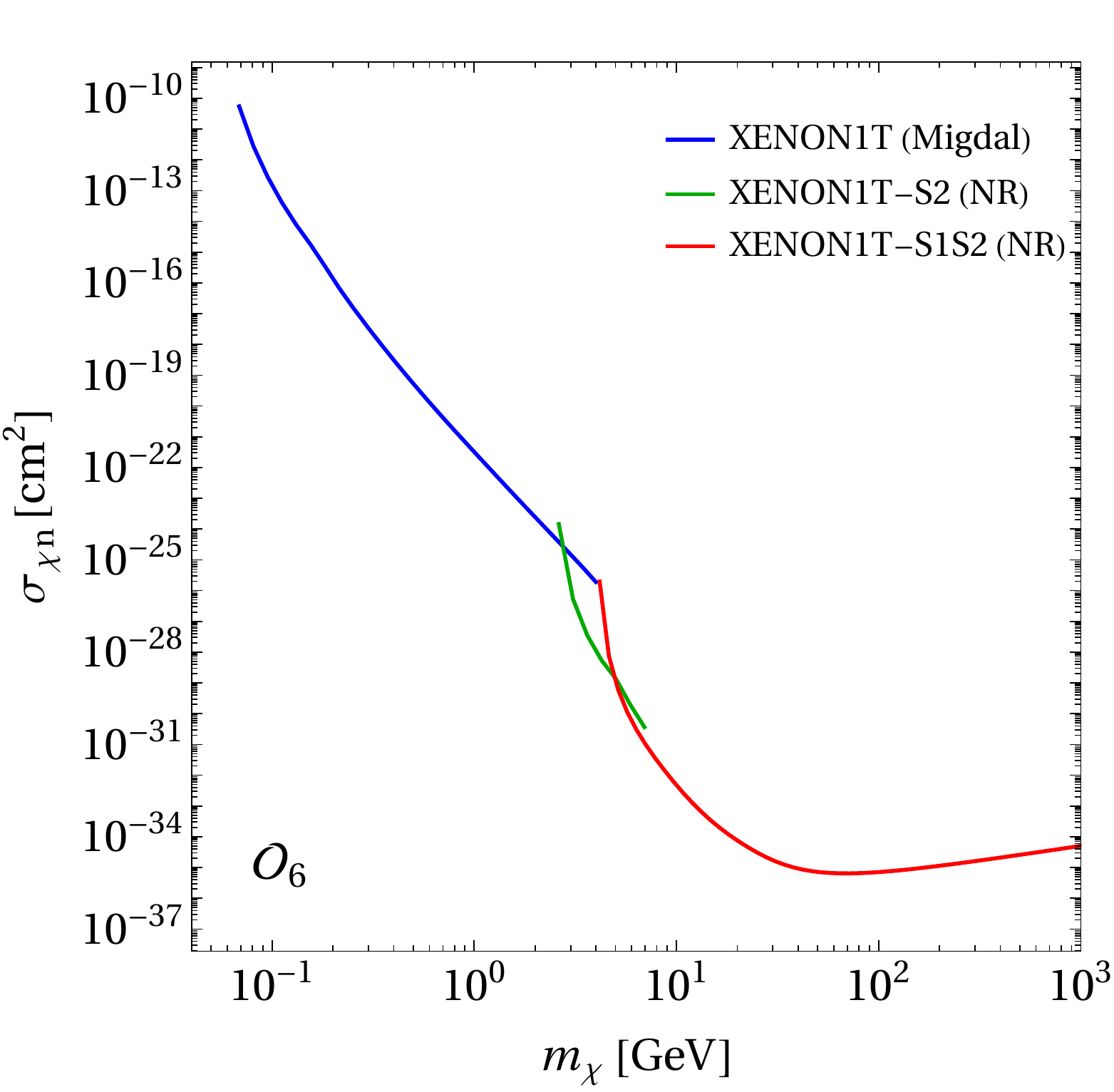} 
&
 \includegraphics[width=5.8cm]{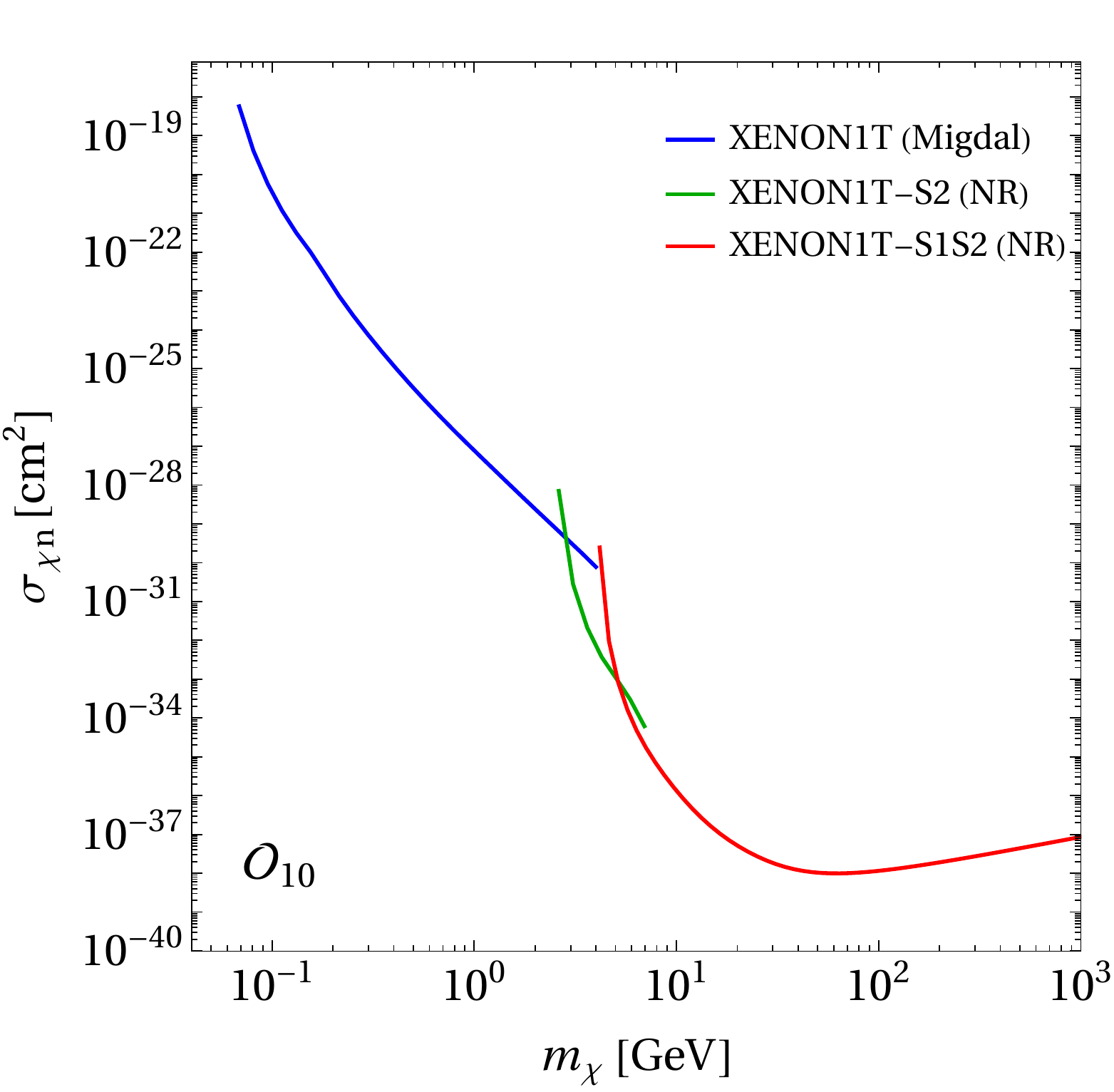}\\
\end{tabular}
\caption{The bounds from XENON1T on the spin- and momentum-dependent operators from the S1S2 NR analysis (red), S2-only NR analysis (green) and the S2-only Migdal analysis (blue), dashed show official limits from the XENON1T collaboration. Bounds from CRESST-III~\cite{Abdelhameed:2019hmk}, Collar~\cite{Collar:2018ydf}, CDMSlite~\cite{Agnese:2017jvy} and CRDM scattering in Borexino~\cite{Bringmann:2018cvk} are also shown for comparison.}
\label{fig:opExclusion}
\end{figure*}

For the spin-dependent operators, we place bounds on the neutron and proton cross sections separately, as shown in Fig.~\ref{fig:opExclusion}. The bounds on each operator follow the same general shape whereby the Migdal effect produces stronger bounds than those from nuclear recoils below around $m_\chi \sim 3$ GeV, falling off sharply for masses below $0.1$ GeV. The Migdal-nuclear bound crossover point is dictated by the detector dependent threshold. Where available we have included the official bounds from XENON1T. The disagreement between our result and the official results is due to the use of different spin structure functions. The similarity in the Migdal effect bounds on each operator (up to a scaling factor) is due to the insensitivity of the shape of the Migdal recoil spectrum to the underlying nuclear recoil operator, as pointed out in section \ref{sec:rates}. For very small dark matter masses the strongest constraints come from cosmic ray dark matter~\cite{Bringmann:2018cvk}, where dark matter is up-scattered to GeV energies by cosmic ray protons and helium. While these constraints, as well as those from protonated scintillators \cite{Collar:2018ydf}, can be applied to all operators, only bounds for $\mathcal{O}_4$ proton cross sections are available. Xenon isotopes, having an even number of protons and thus a small proton spin expectation value, are ill-suited to constraining the spin-dependent proton cross sections. Better bounds from the Migdal effect on the proton spin-dependent cross sections could be obtained from a proton spin-rich target such as fluorine. However, presently the best proton spin-dependent bounds are from PICO-60~\cite{Amole:2017dex}, which is insensitive to electronic recoils and therefore also to the Migdal effect.

\section{Coherent Elastic Neutrino-Nucleus Scattering}
\label{sec:neutrino}

The formalism developed for the Migdal effect and photon emission during dark matter-nucleus scattering can also be applied to the case of coherent elastic neutrino-nucleus scattering \cns\footnote{Recently bremsstrahlung in the \cns${}$ process was discussed as a possible means of determining the neutrino mass by examining the photon endpoint~\cite{Millar:2018hkv}.}. In this case, the differential cross-section is already known to be dominantly spin-independent, having been calculated in the 1970s \cite{Freedman:1973yd}. The first experimental evidence of the process was announced by the COHERENT collaboration in 2017 \cite{Akimov:2017ade}. Following the discussion of Section \ref{sec:migdal_brem}
, we utilize Eqs.(\ref{eq:migdal_rate}) and~(\ref{eq:PradlerBrem}) with the \cns~cross section
\bea
\left(\frac{d\sigma}{dE_R}\right)_{\rm{CE\nu NS}}\!\! = \frac{G_F^2m_T}{4\pi}Q_{V}^2\left(1 - \frac{m_T E_R}{2E_\nu^2}\right)\!F^2(q^2)
\eea
In this expression, $G_F$ is the Fermi constant, $E_\nu$ is the incident neutrino energy, $F(q^2)$ is the nuclear form factor which encodes the momentum dependence, or equivalently the loss of coherence. In what follows, we will adopt the standard Helm form factor for $F(q^2)$ \cite{Helm:1956zz}. The quantity $Q_V$ is the vector charge of the nucleus, and in the \cns process, it depends dominantly on the neutron number, $N$: $Q_{V} = N - (1-4\sin^2\theta_W)Z\simeq N - (.08)Z$. As discussed below, the kinematic limits for both the Migdal effect and bremsstrahlung  must be adjusted for \cns as compared to the massive dark matter case, due to the neutrino's relativistic nature.

As first outlined in \cite{Ibe:2017yqa}, the \cns process can produce ionization through the Migdal effect~\footnote{The \cns cross section given in Eq.~(118) of \cite{Ibe:2017yqa} is a factor of 2 smaller than the formula presented here. Our Eq.~(17) is in agreement with other literature on the topic, see for example~\cite{Cerdeno:2016sfi}.}. The differential cross-section becomes
\bea
\frac{d\sigma}{dE_R} =&& \frac{G_F^2}{4\pi}Q_V^2m_T
\\
\nonumber
&&\times\left(1 - \frac{m_TE_R - E_{EM}^2}{2E_\nu^2}\right)F^2(q^2)\times|Z_{{\rm{ion}}}(q_e)|^2
\eea
and the kinematic endpoints are also slightly altered due to the electronic energy injection. The nuclear recoil now lies in the range (see Appendix \ref{app:nuMigdalKinematics} for the details)
\bea
\frac{\left(E_e + E_{n\ell}\right)^2}{2m_T} < E_R < \frac{\left(2E_\nu - (E_e + E_{n\ell})\right)^2}{2(m_T + 2E_\nu)}
\eea

\begin{figure*}[htb]
\begin{tabular}{cc}
\includegraphics[width=7cm]{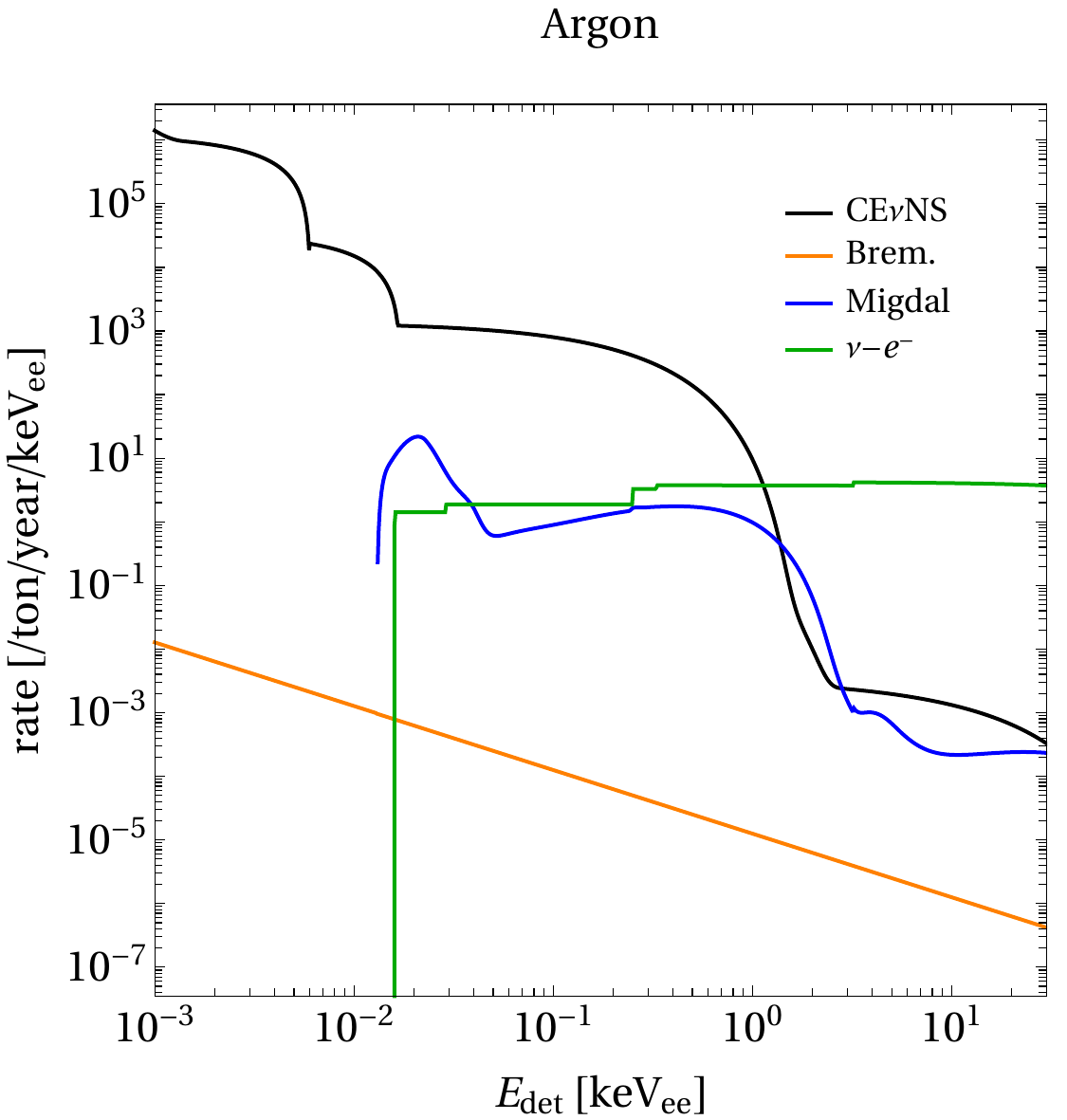} 
& \includegraphics[width=7cm]{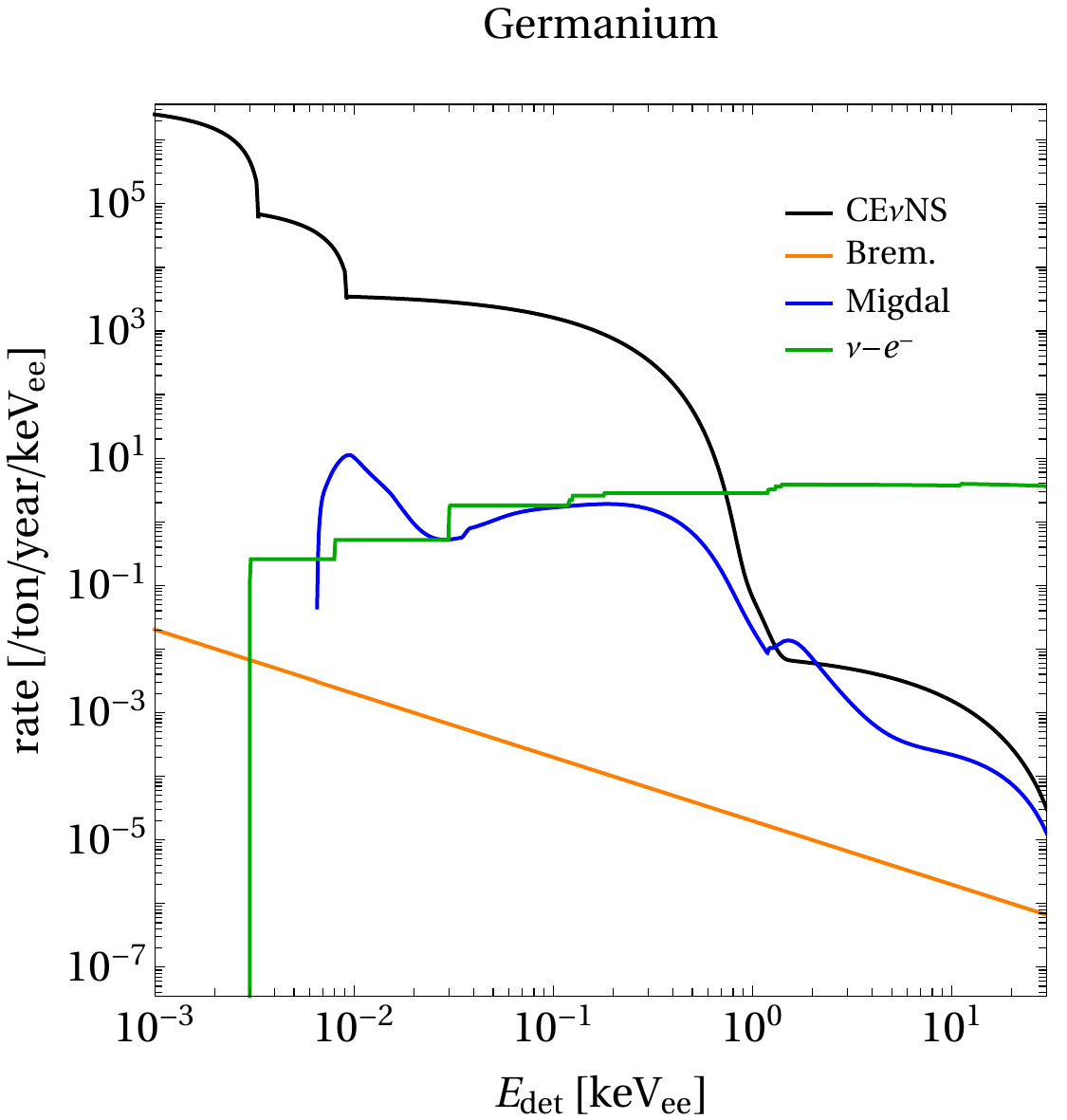}\\
\includegraphics[width=7cm]{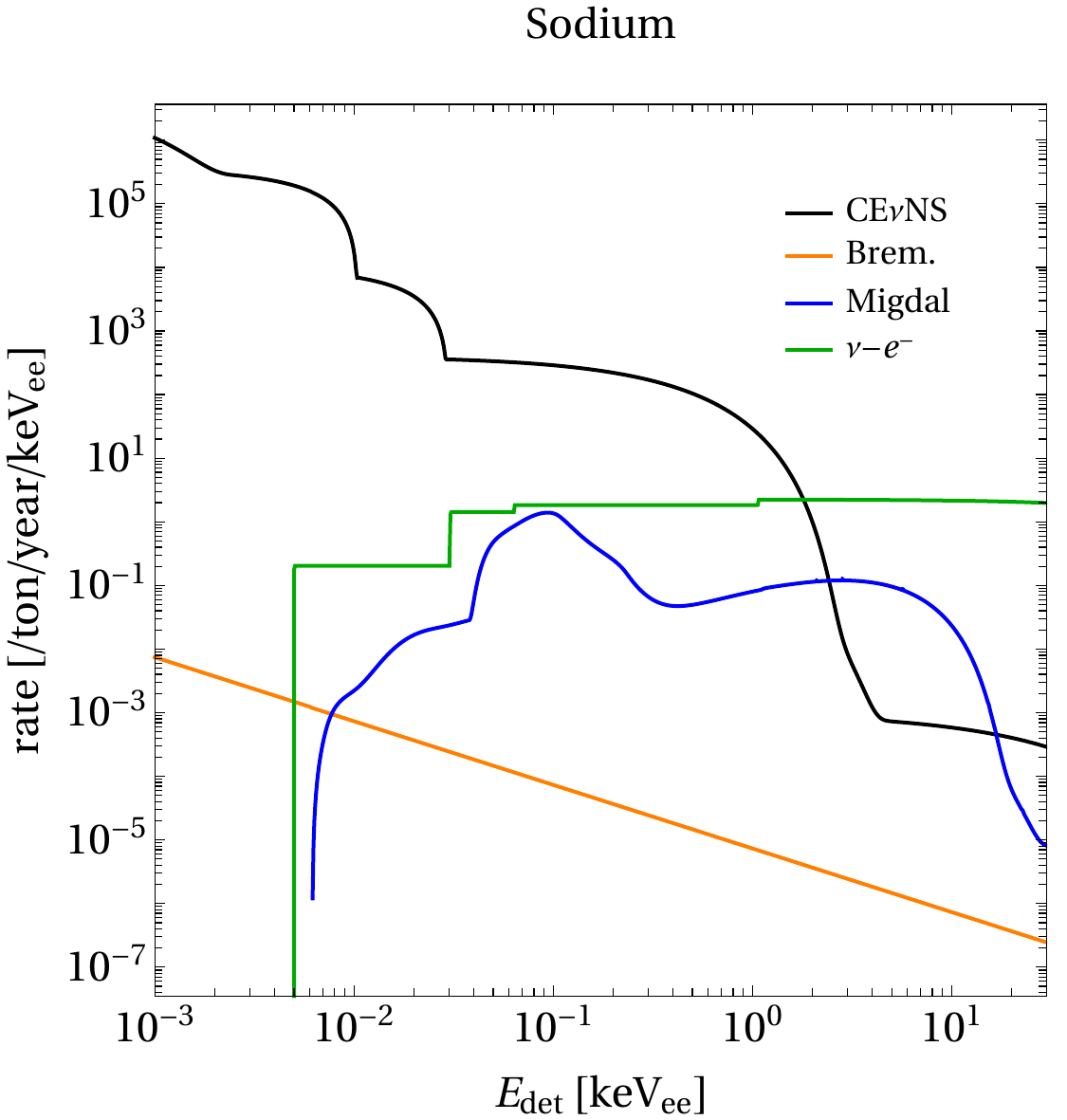} 
& \includegraphics[width=7cm]{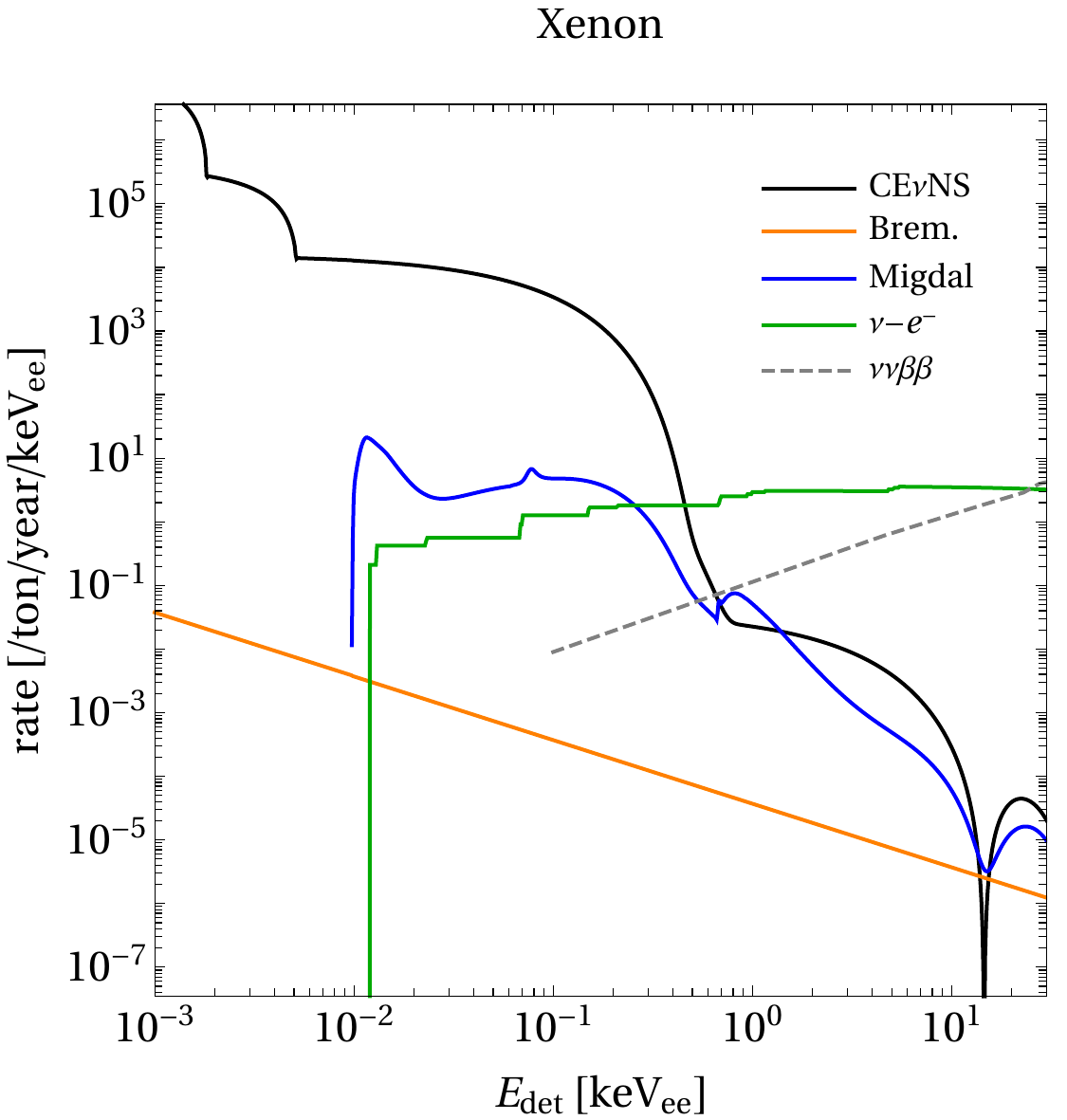}\\
\end{tabular}
\caption{The total rate for the Migdal effect and bremsstrahlung induced by coherent neutrino-nucleus scattering from solar and atmospheric neutrinos for the four targets considered: argon, germanium, sodium and xenon. For comparison, the rate for \cns (electronic recoils) is depicted by the black (green) line.}
\label{fig:bremNu}
\end{figure*}

For bremsstrahlung, the double differential cross-section, $d^2\sigma/d\omega dE_R$, needs to be integrated between the following nuclear recoil energy limits, as derived in Appendix \ref{app:nuBremKinematics}
\be
E_{R,\rm{min}} = \frac{2\omega^2}{m_T - 2\omega},\,\,\, E_{R,\rm{max}} = \frac{2E_{\nu}^2}{m_T + 2E_{\nu}}.
\ee
Finally, the rate for the CE$\nu$NS bremsstrahlung process is
\bea
\frac{dR_{\nu\gamma}}{d\omega} = N_T\sum_{\alpha}\int \left(\frac{d\Phi}{dE_\nu}\right)_{\alpha}\left(\frac{d\sigma}{d\omega}\right)_{\rm{CE\nu NS}}dE_\nu
\eea
where $d\Phi_\alpha/dE_\nu$ is the incident neutrino flux, $\alpha$ denotes the neutrino flux source, and $E_{\nu,\mathrm{min}}=\omega$. In this work we examine the solar neutrino   fluxes: $pp$, $pep$, $^8$B and $^7$Be, as well as the flux of atmospheric neutrinos~\cite{Billard:2013qya}. The result of our calculation is given in Fig.~\ref{fig:bremNu}. Examination of the results clearly shows that the Migdal effect has a rate (due to solar neutrinos) that is comparable to that of the atmospheric neutrinos fluxes in the few keV energy range, while the bremsstrahlung effect is sub-dominant to the total nuclear recoil rate over the entirety of the energy range.
 These results demonstrate that the Migdal effect could play an interesting role for future G3 dark matter direct detection experiments with sensitivity to atmospheric neutrinos, while the effects of bremsstrahlung can be neglected.

\section{Summary}
\label{sec:summary}

Experimental efforts in the field of dark matter direct detection have made tremendous progress in probing the WIMP parameter space, both to lower masses and to lower cross sections. However, low threshold detectors come at the expense of detector size. Thus, simultaneously probing lower masses and lower cross sections is a challenge. Both the Migdal effect and bremsstrahlung from interactions with very light WIMPs provide pathways for experiments to probe below their nominal nuclear recoil thresholds. This extends the experimental reach in the WIMP parameter space. In this paper we have recast the Migdal effect and bremsstrahlung rates within an effective field theory context capable of including a general set of spin- and momentum-dependent interactions of the WIMP and nucleons. We extend and generalize the analyses for the Migdal effect in~\cite{Ibe:2017yqa,Dolan:2017xbu} and the bremsstrahlung work of~\cite{Kouvaris:2016afs}. 

We calculated the rates associated with the Migdal effect and photon bremsstrahlung for a representative set of effective field theory operators and detector materials. We found that across all combinations of interactions and targets considered, the Migdal rate dominates over the bremsstrahlung rate, and can provide the dominant signal for direct detection experiments for sub-keV energy deposits. A signal at these low energies would correspond to dark matter masses in the sub-GeV range. We used the Migdal rates to place new experimental constraints on the EFT operators for low mass WIMPs.

Finally, we applied the Migdal and bremsstrahlung formulations to the case of \cns, which will soon become an irreducible background in dark matter experiments. We calculated the rates for these processes induced by solar and atmospheric neutrinos. We demonstrated that the contribution from bremsstrahlung is sub-dominant to the elastic rate, and can thus be safely ignored.  However, the Migdal effect from ${}^8$B solar neutrinos can provide a competitive or leading signal in the region typically expected to be dominated by elastic nuclear recoils from atmospheric neutrinos. Thus, more careful study of the Migdal effect in this energy range should be undertaken.


\section*{Acknowledgements}
The authors would like to thank Wick Haxton, Jeremy Holt, and Sergei Maydanyuk for helpful discussion regarding nuclear physics aspects of this project. The authors also are grateful to Alex Millar, Georg Raffelft, Leo Stodolsky, and Edoardo Vitagliano for correctly identifying substantive issues with an earlier version of this work.  NFB was supported by the Australian Research Council. JBD acknowledges support from the National Science Foundation under Grant No. NSF PHY-1820801. JBD would also like to thank the Mitchell Institute for Fundamental Physics and Astronomy at Texas A\&M University, the University of Melbourne, and the Kavli Institute for Theoretical Physics for their generous hospitality during various stages of this work. JLN thanks the University of Melbourne for their hospitality during portions of this work. TJW acknowledges ``investigador convidat" (UV-INV-EPC 2018) support from the U. Valencia. This research was also supported in part by the National Science Foundation under Grant No. NSF PHY-1748958 (JBD), and US DoE Grant DE-SC-001198 (TJW).


\appendix


\section{Rates for $m_{\chi} = 2$ GeV and $m_{\chi} = 0.5$ GeV}
\label{app:2and5rates}

\begin{figure*}[hbt]
\begin{tabular}{ccc}
\includegraphics[width=5.8cm]{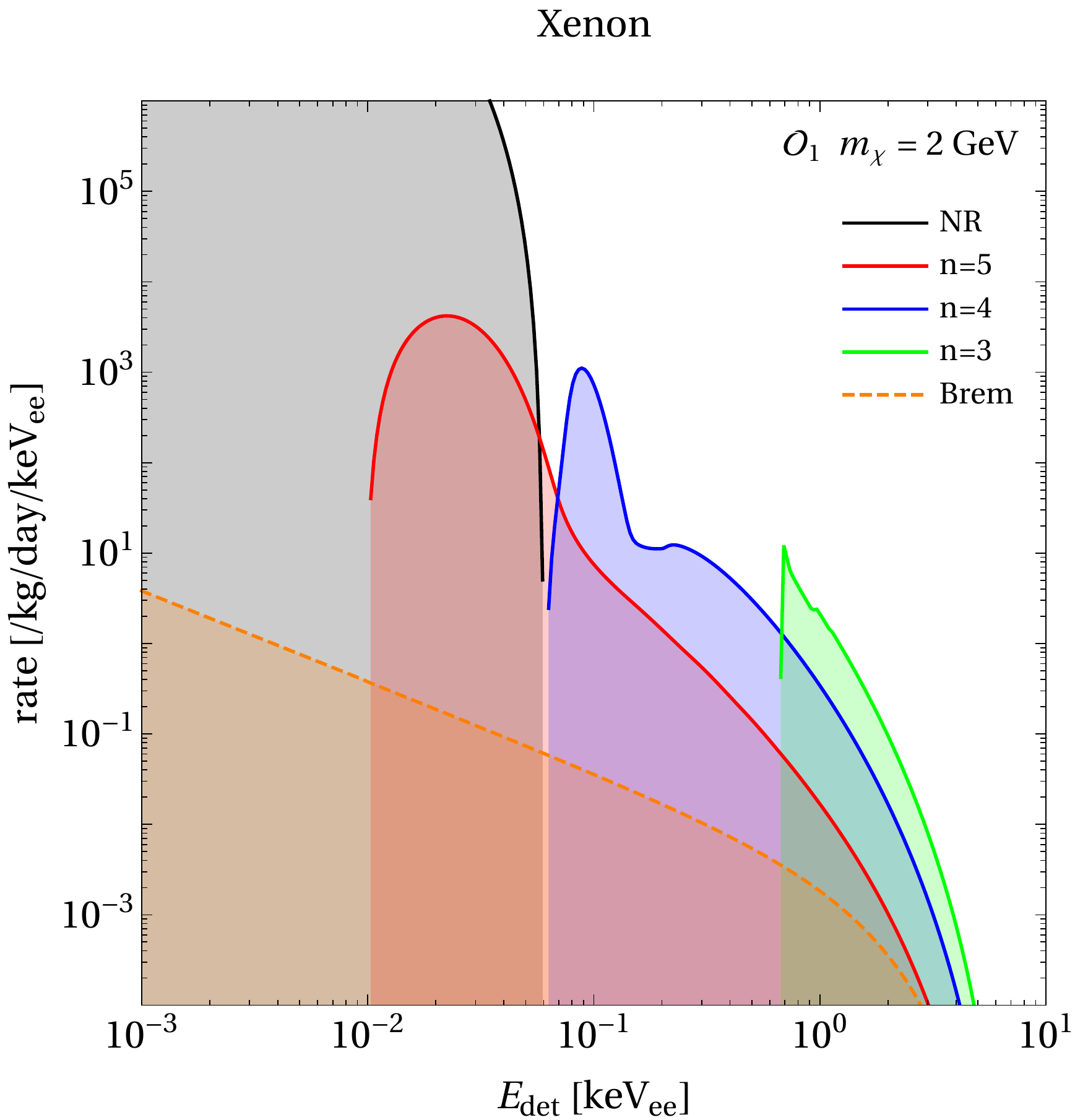} 
&
\includegraphics[width=5.8cm]{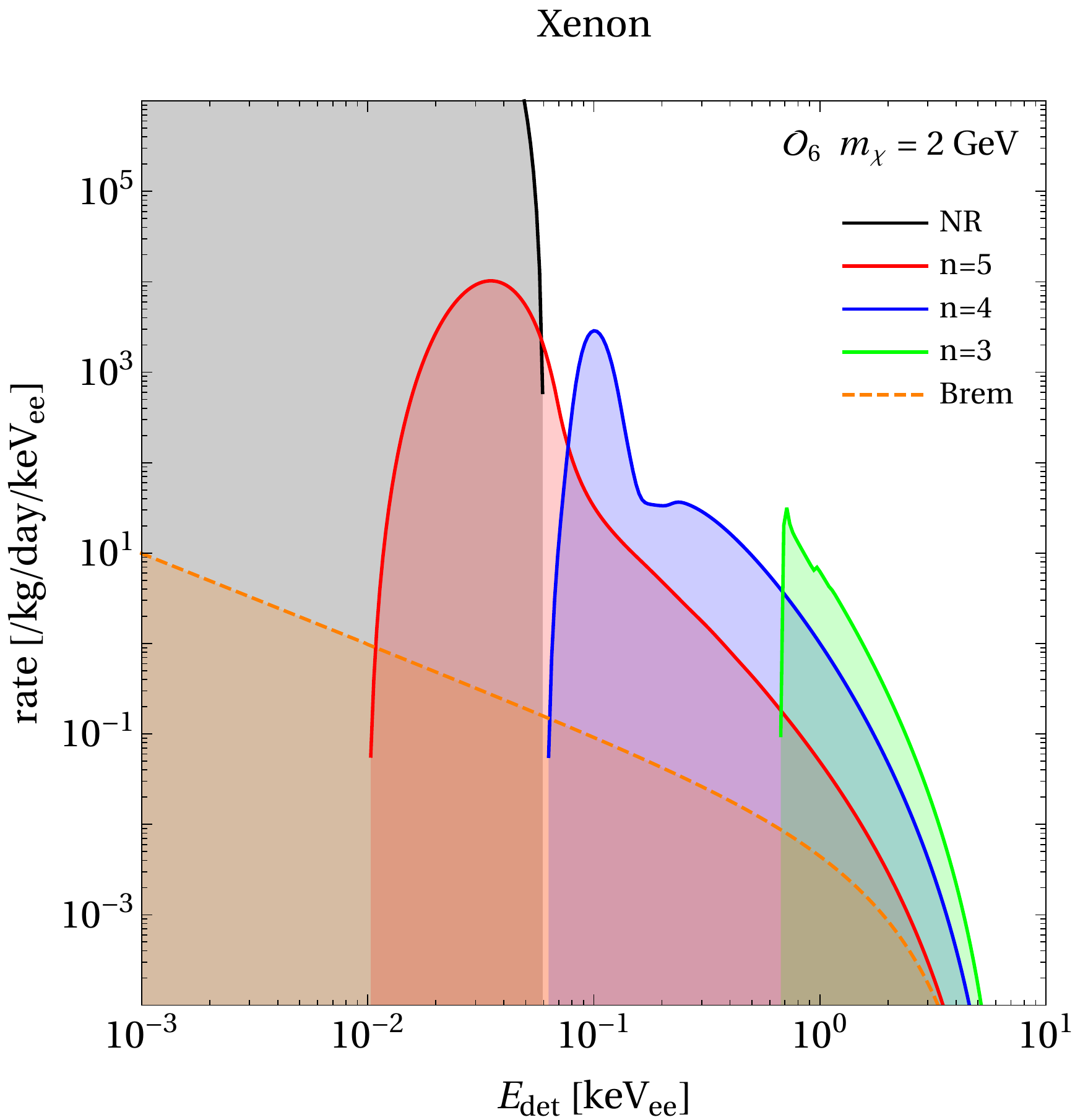}
&
\includegraphics[width=5.8cm]{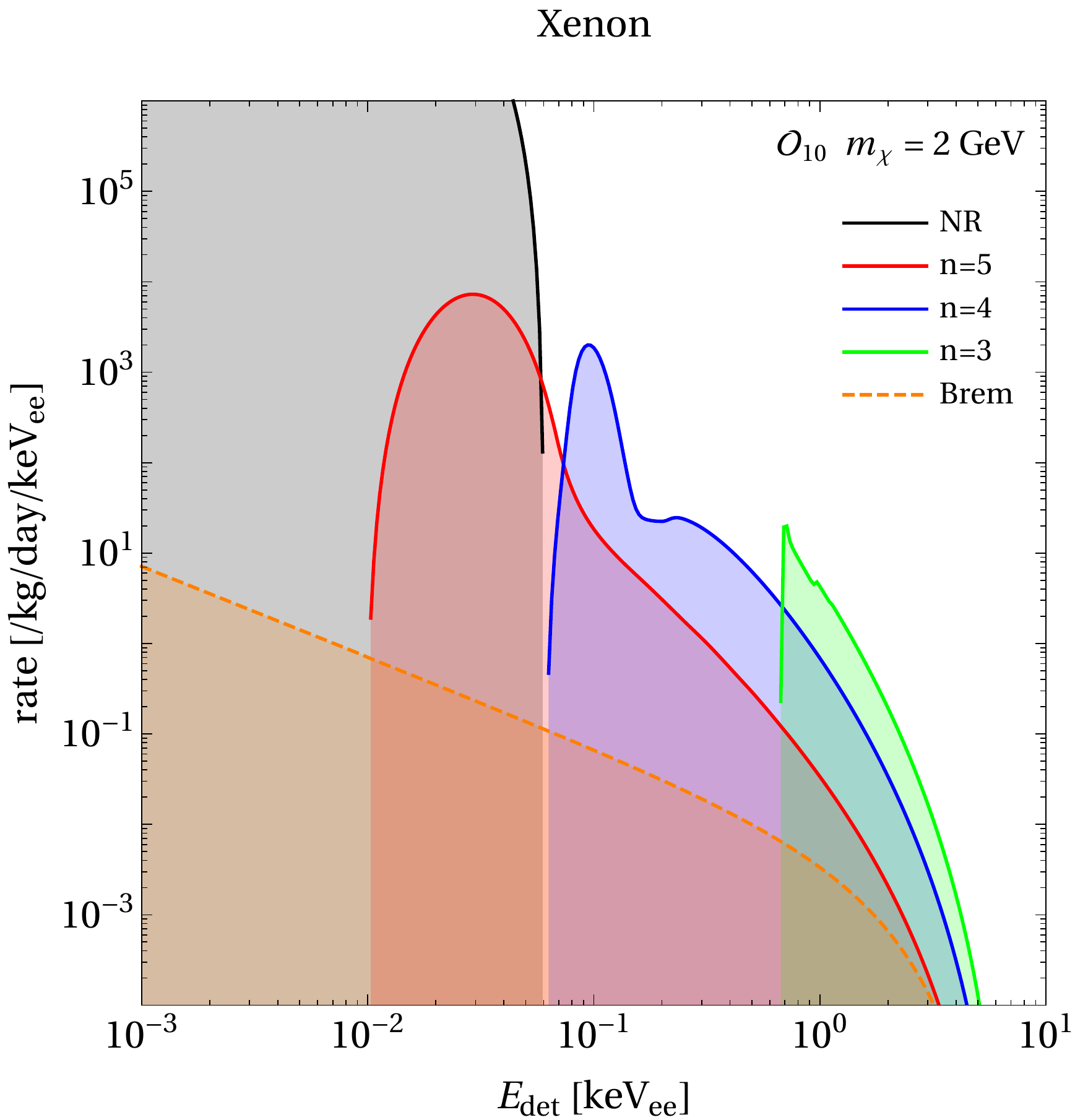}\\
\includegraphics[width=5.8cm]{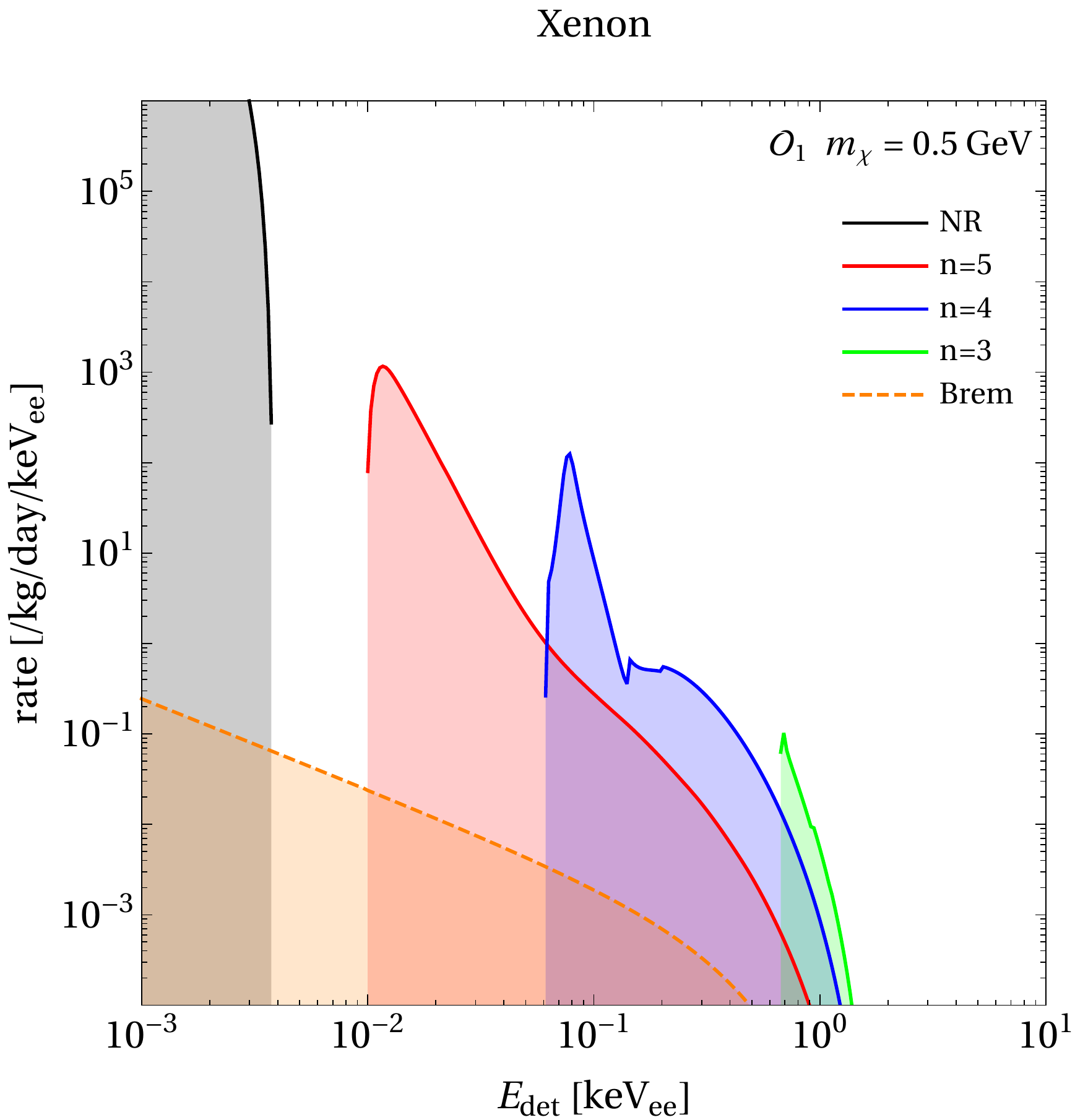} 
&
\includegraphics[width=5.8cm]{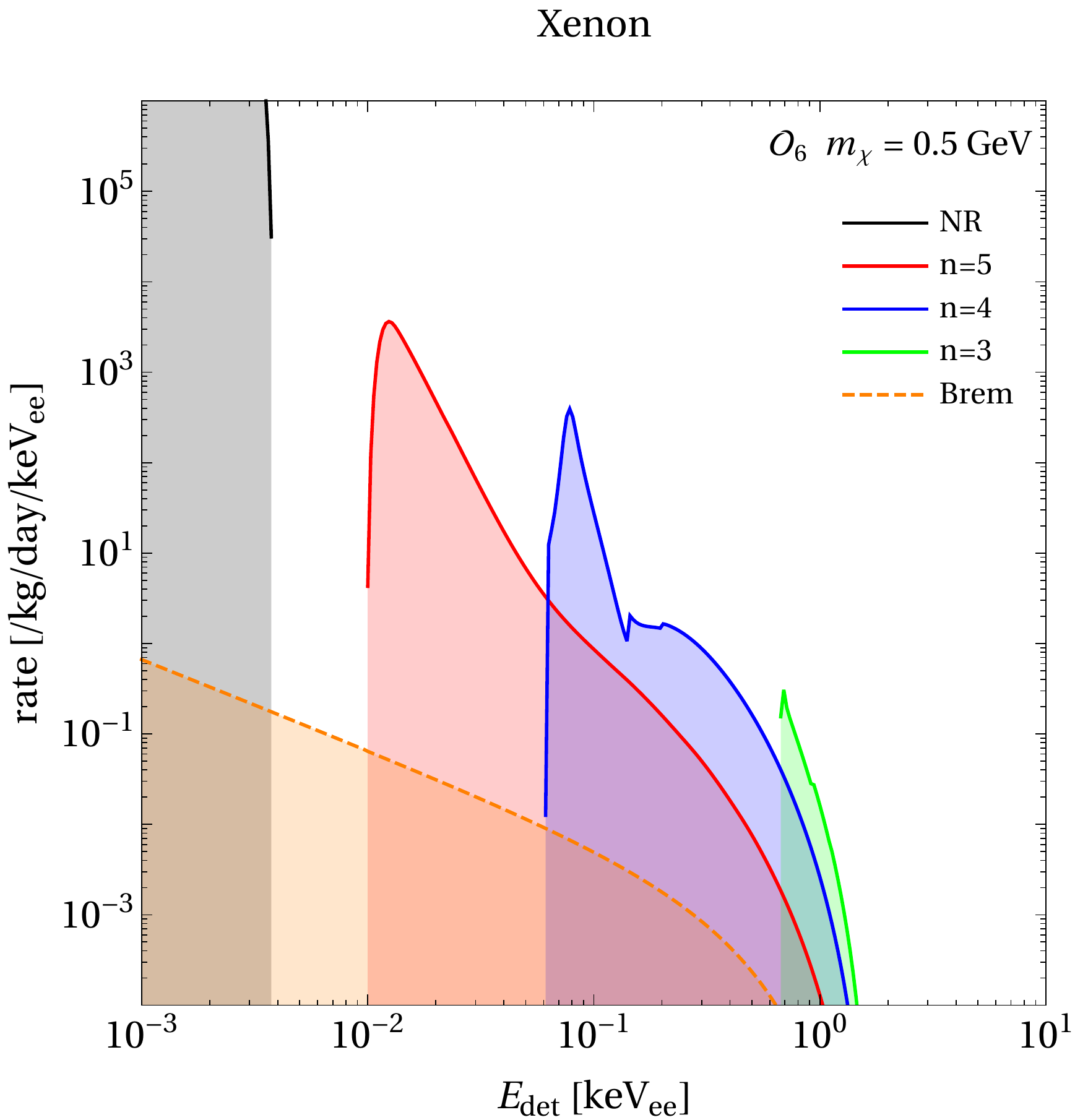}
&
\includegraphics[width=5.8cm]{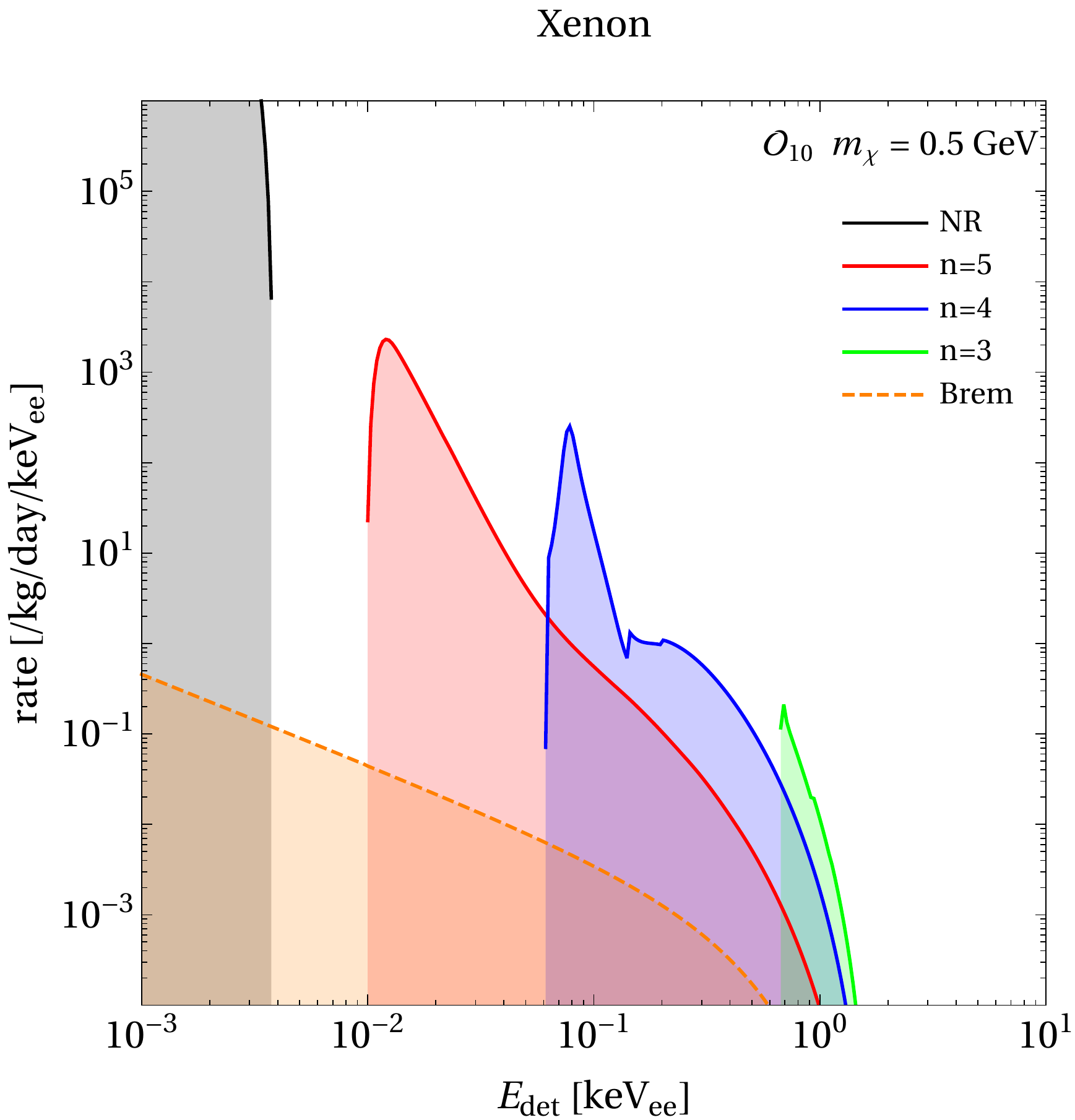}\\
\end{tabular}
\caption{The rate for the Migdal effect in xenon (depicted by the colored lines associated with various atomic energy levels denoted by $n$) and bremsstrahlung (dashed orange line) induced by a 2 GeV (top) and 0.5 GeV (bottom) mass dark matter particle interacting with the nucleus through the operators under consideration (the rate for $\mathcal{O}_4$ is omitted here since it has the same energy dependence as $\mathcal{O}_1$).}
\label{fig:otherMasses}
\end{figure*}

We present the rates (for Xenon) from elastic nuclear recoils, the Migdal effect, and photon bremsstrahlung for the operators $\mathcal{O}_1$, $\mathcal{O}_6$, and $\mathcal{O}_{10}$ and for dark matter masses of $m_{\chi} = 0.5$ GeV and $m_{\chi} = 2$ GeV in Fig.~\ref{fig:otherMasses}. These figures clearly illustrate that our principal conclusions are same as those for $m_{\chi} = 1$ GeV case considered above.

\section{Phase Space}
\label{app:phaseSpace}

Here we provide a derivation of the phase space for the $2\rightarrow3$ scattering processes. We begin with the four-momenta in the lab frame, where the $k$/$k'$ momenta now denote that of the incoming/outgoing nucleus respectively:
\bea
p &=& (E_\chi, \vec{p}),\\
k &=& (m_T,0),\\
p' &=& (E_\chi',\vec{p'}),\\
k'&=& (m_T+E_R,\vec{k'}),\\
\omega &=& (\omega, \vec{\omega}).
\eea
The differential cross-section is given by
\begin{widetext}
\bea
d\sigma &=& \frac{1}{2E_\chi 2E_k v}\frac{d^3\vec{p'}}{(2\pi)^32E_\chi'}\frac{d^3\vec{k'}}{(2\pi)^32E_{k'}}\frac{d^3\vec{\omega}}{(2\pi)^32\omega}(2\pi)^4|\mathcal{M}|^2_\twothree\delta^4(p+k-p'-k'-\omega)\\
&=& \frac{1}{2^{10}\pi^5E_\chi E_k v}\frac{d^3\vec{p'}}{E_\chi'}\frac{d^3\vec{k'}}{E_{k'}}d\omega d\Omega_{\omega}\omega|\mathcal{M}|^2_\twothree\delta^4(p+k-p'-k'-\omega).
\eea
We can use the spatial delta-function to do the $\vec{p'}$ integrations
\bea
\frac{d\sigma}{d\omega} &=& \frac{1}{2^{10}\pi^5E_\chi E_k v}\frac{1}{E_\chi'}\frac{d^3\vec{k'}}{E_{k'}} d\Omega_{\omega}\omega|\mathcal{M}|^2_\twothree\delta(E_\chi + m_T - E_\chi' - E_{k'} - \omega)\\
&=& \frac{1}{2^{10}\pi^5E_\chi E_k v}\frac{1}{E_\chi'}\frac{d\vk \vk^2}{E_{k'}}d\Omega_{k'} d\Omega_{\omega}\omega|\mathcal{M}|^2_\twothree\delta(E_\chi + m_T - E_\chi' - E_{k'} - \omega).
\eea
Again we will change variables
\bea
\frac{d\sigma}{dE_R} = \frac{d\Omega_{k'}}{dE_R}\frac{d\sigma}{d\Omega_{k'}} = -2\pi \frac{d\cos\theta_{k'}}{dE_R}\frac{d\sigma}{d\Omega_{k'}}.
\eea
In the current case this yields
\bea
\frac{d^2\sigma}{dE_R d\omega} &=& -2\pi\frac{d\cos\theta_{k'}}{dE_R}\frac{d^2\sigma}{d\omega d\Omega_{k'}}\nn
&=&-2\pi\frac{d\cos\theta_{k'}}{dE_R}\frac{1}{2^{10}\pi^5E_\chi E_k v}\frac{1}{E_\chi'}\frac{d\vk \vk^2}{E_{k'}} d\Omega_{\omega}\omega\nn
& &\times \qquad |\mathcal{M}|^2_\twothree\delta(E_\chi + m_T - E_\chi' - E_{k'} - \omega).
\eea
In order to find $\cos\theta_{k'}$ as a function of $E_R$, we need the momentum relation
\bea
p - k' - \omega = p' - k,
\eea
which leads to
\bea
p\cdot k' + p\cdot \omega - k'\cdot\omega = p'\cdot k.
\eea
Using the kinematic relations, we find
\bea
E_{\chi}(m_T+E_R) - |\vec{p}|\vkp\cos\theta_{k'} + E_\chi \omega - |\vec{p}|\omega \cos\theta_{\omega} - \omega(E_R + m_T) + \vkp\omega\cos\theta_{k'\omega} = E_\chi' m_T,
\eea
which, together with  $E_\chi + m_T = E_\chi' + m_T + E_R + \omega$, results in
\bea
\cos\theta_{k'} = \frac{E_R(m_T+E_\chi -\omega)}{|\vec{p}|\vkp} + \frac{\omega(E_\chi - |\vec{p}|\cos\theta_\omega)}{|\vec{p}|\vkp} + \frac{\omega\cos\theta_{k'\omega}}{|\vec{p}|},
\eea
and hence
\bea
\frac{d\cos\theta_{k'}}{dE_R} &=& \frac{m_T+E_\chi - \omega}{2|\vec{p}|\vkp} - \frac{2m_T\omega(E_\chi - |\vec{p}|\cos\theta_\omega)}{2|\vec{p}|\vkp^3}\\
&=& \frac{m_T}{|\vec{p}|\vkp^3}\left(E_R(m_T+E_\chi - \omega) -\omega(E_\chi - |\vec{p}|\cos\theta_\omega)\right).
\eea
For the delta function we use 
\bea
\delta(E_\chi + m_T - E_\chi' - E_{k'} - \omega) &=& \delta\left(E_\chi + m_T - \omega \right. \\\nonumber
&& - (m_\chi^2 + |\vec{p}|^2 + |\vec{\omega}|^2+\vkp^2 - 2|\vec{p}|\omega\cos\theta_\omega - 2|\vec{p}|\vkp\cos\theta_{k'} + 2\vkp\omega\cos\theta_{\omega k'})^{1/2} \\\nonumber 
&& \left. - (m_T^2 + \vkp^2)^{1/2}\right)\\
&=&-\delta(\vkp - X)\frac{E_{k'}E_\chi'}{\vkp(E_\chi + m_T - \omega) + E_{k'}(-|\vec{p}|\cos\theta_{k'} + \omega\cos\theta_{\omega k'})}.
\eea
Next we use the momentum relation,
\bea
-|\vec{p}|\cos\theta_{k'} + \omega\cos\theta_{\omega k'} = \frac{E_R(\omega - m_T - E_\chi) - \omega(E_\chi - |\vec{p}|\cos\theta_\omega)}{\vkp},
\eea
to find
\bea
\delta(E_\chi + m_T - E_\chi' - E_{k'} - \omega) = -\delta(\vkp - X)\frac{E_{k'}E_\chi'\vkp}{\vkp^2(E_\chi + m_T - \omega) + (m_T+E_R)(E_R(\omega - m_T - E_\chi) - \omega(E_\chi - |\vec{p}|\cos\theta_\omega)}.\nn
\eea
Using the good approximation $m_T+E_R \simeq m_T$, we have
\bea
\delta(E_\chi + m_T - E_\chi' - E_{k'} - \omega) = -\delta(\vec{k'} - \vec{X})\frac{E_{k'}E_\chi'\vkp}{m_T E_R(E_\chi + m_T - \omega) -m_T\omega(E_\chi - |\vec{p}|\cos\theta_\omega)}.
\eea
We then obtain the simplification
\bea
-2\pi\frac{d\cos \theta_{k'}}{dE_R}\delta(E_\chi + m_T - E_\chi' - E_{k'} - \omega) = \frac{2\pi}{|\vec{p}|\vkp^2}\delta(\vk - X).
\eea
\end{widetext}
Using the delta function to do the $d\vkp$ integration, we arrive at
\bea
\frac{d^2\sigma}{dE_Rd\omega} = \frac{1}{2^{9}\pi^4E_\chi m_T v}\frac{\omega}{|\vec{p}|}d\Omega_{\omega} |\mathcal{M}|^2_\twothree .
\eea
The $2\rightarrow3$ amplitude factorizes to produce the double differential cross section~\cite{Kouvaris:2016afs}
\bea
\frac{d^2\sigma}{dE_Rd\omega} &=& \frac{e^2}{2^4\pi^3}\frac{q^2\cos^2\theta_{\omega}}{\omega^2m_N^2}d\Omega_{\omega}\left(\frac{d\sigma^{(p)}}{dE_R}\right)_\twotwo\nn
&=& \frac{4\alpha}{3\pi}\frac{m_T E_R}{m_N^2\omega}\left(\frac{d\sigma^{(p)}}{dE_R}\right)_\twotwo,
\eea
where the final line includes an extra factor of 2 from summing over both polarizations.


\section{Kinematics}
\label{app:kinematics}

\subsection{Dark matter scattering and the Migdal effect}

We begin in the center-of-mass frame where the incident dark matter momentum and target atom momentum are related by
\bea
\mchi\vcm + m_T\vtm = 0.
\eea
In the non-relativistic framework, the boost to the lab frame, where the target is at rest, is provided by the relation
\bea
\vcm = \vcl + \vtm,
\eea
or, given in terms of magnitudes,
\bea
v_{\chi,{\rm{COM}}} = v_{\chi,{\rm{Lab}}} - v_{T,{\rm{CoM}}}.
\eea

The relation between the dark matter momentum in the center-of-mass frame and the incident dark matter velocity in the laboratory frame is
\bea
\mchi v_{\chi,{\rm{CoM}}} = \mu_T v_{\chi,{\rm{Lab}}}.
\eea
Now we can treat the scattering as an incoming dark matter particle of momentum $p_\chi$ colliding with a target at rest in the lab frame represented by momentum $k$, and producing dark matter of momentum $p_\chi'$ and target atom of momentum $k'$ scattered at a lab angle of $\theta$ with nuclear recoil energy, $E_R$. We represent these by
\bea
p_\chi &=& (E_{\chi},\vec{p}_{\chi}),
\\
k &=& (m_T + E_{e,i},0),
\\
p_\chi' &=& (E_{\chi}',\vec{p}_{\chi}'),
\\
k' &=& (m_T + E_{e,f} + E_R,\vec{k}'),
\eea
where the energy of the electron cloud in the initial and final states are represented by $E_{e,i}$ and $E_{e,f}$, respectively. 
Energy and three-momentum conservation lead to
\bea
E_{\chi} + m_T &=&  E_{\chi}' + m_T + E_R + E_{{\rm{EM}}},
\\
\vec{p}_{\chi} &=& \vec{p}_{\chi}' + \vec{k}',
\eea
where $E_{{\rm{EM}}} = E_{e,f} - E_{e,i}$. Using $\vec{p}_{\chi}\cdot\vec{k}' = |\vec{p}_{\chi}||\vec{k}'|\cos\theta$ and the usual energy-momentum relation $E^2 = m^2 + |\vec{p}|^2$, we find
\bea
2m_TE_R - 2m_\chi v\cos\theta\sqrt{2m_TE_R} = -2E_\chi E_R - 2E_\chi E_{{\rm{EM}}},\nn
\eea
where we have dropped terms quadratic in $E_R$ and $E_{{\rm{EM}}}$ \emph{relative} to $E_\chi E_R$ and $E_\chi E_{{\rm{EM}}}$, and used that the momentum exchanged is $|\vec{q}| \simeq \sqrt{2m_TE_R}$, which is corrected at order $(E_{{\rm{EM}}})^2/(2m_TE_R)$. We will also use the well founded approximation $E_\chi \simeq m_\chi \simeq E_{\chi}'$.

We now have the quadratic equation for the recoil energy
\bea\label{MigdalER}
E_R^2\left(\frac{m_T}{\mu_T}\right)^2 &+& E_R\left[2E_{{\rm{EM}}}\left(\frac{m_T}{\mu_T}\right) -2m_Tv^2\cos^2\theta\right]\nn 
&+& E_{{\rm{EM}}}^2 = 0.
\eea
The solutions are
\bea
\label{eq:MigdalKin}
E_R &=& \frac{\mu_T^2 v^2\cos^2\theta }{m_T}\left[\left(1 - \frac{E_{{\rm{EM}}}}{\mu_T v^2\cos^2\theta}\right)\right.\\
&\pm& \left.\sqrt{ 1 - \frac{2E_{{\rm{EM}}}}{\mu_T v^2\cos^2\theta}}\right].\nonumber
\eea
It should come as no surprise that this relation is the same as that for the recoil energy in the case of inelastic dark matter with the change in electron cloud energy substituted for the inelastic mass-splitting parameter $\delta$ (this was also pointed out in \cite{Dolan:2017xbu}.  See, for example, the non-numbered first equation in Sec.~2 of \cite{Bramante:2016rdh}). 

From the above expression for $E_{\rm{R}}$, we see that the minimum speed for the dark matter to produce a nuclear recoil of energy $E_R$ and the accompanying electromagnetic energy injection from the Migdal effect, $E_{{\rm{EM}}}$, as given in \cite{Ibe:2017yqa}
\bea
v_{\rm{min}} = \frac{m_TE_R + \mu_TE_{\rm{EM}}}{\mu_T\sqrt{2m_TE_R}}.
\eea
We also see the maximum values for the recoil and electronic energies are set by the maximum dark matter speed, \cite{Ibe:2017yqa,Dolan:2017xbu} 
\bea
E_{R,{\rm{max}}} &=& \frac{2\mu_T^2v_{{\rm{max}}}^2}{m_T},
\\
E_{{\rm{EM}},{\rm{max}}} &=& \frac{\mu_Tv_{{\rm{max}}}^2}{2}.
\eea
For $E_{R,{\rm{max}}}$, we set $E_{\rm{EM}}=0$ and for $E_{{\rm{EM}},{\rm{max}}}$, we set $E_{R}=E_{{\rm{EM}},{\rm{max}}}$ in (\ref{MigdalER}), respectively.\\

\subsection{Dark matter scattering and photon bremsstrahlung}

The four-momenta in the lab frame are
\bea
p &=& (m_\chi + \frac{m_\chi v^2}{2}, m_\chi \vec{v}),\\
k &=& (m_T, 0),\\
p' &=& (m_\chi + \frac{m_\chi v'^2}{2},m_\chi \vec{v}'),\\
k' &=& (m_T + E_R, m_T \vec{v}'_k),\\
\omega &=& (\omega,\vec{\omega}),
\eea
where the recoil energy and recoil momentum are
\bea
E_R &=& \frac{m_T|\vec{v}'_k|^2}{2}\\ m_T|\vec{v}'_k| &=& \sqrt{2m_TE_R}.
\eea
From energy and momentum conservation we have
\bea
E_R &=& \frac{m_\chi v^2}{2} - \frac{m_\chi v'^2}{2} - \omega,\\
m_\chi \vec{v}' &=& m_\chi \vec{v} - m_T\vec{v}_{k'} - \vec{\omega}.
\eea
Squaring the last equation, and substituting it into the equation for $E_R$, we find
\bea
E_R &&= -\frac{m_T}{m_\chi}E_R + v\cos\theta_{pk'}\sqrt{2m_T E_R}\nn 
&&+\omega\left(v\cos\theta_{p\omega} -\frac{m_T}{m_\chi}v_k'\cos\theta_{k'\omega} - \frac{\omega}{2m_\chi}-1\right),\nn
\eea
or
\bea
E_R\left(\frac{m_T}{\mu}\right) &=& \sqrt{2m_T E_R}\left(v\cos\theta_{pk'} -\frac{\omega}{m_\chi}\cos\theta_{k'\omega}\right) \nn 
&&+ \omega\left(v\cos\theta_{p\omega}-\frac{\omega}{2m\chi}-1\right),
\eea
where we have adopted the notation $\theta_{ij}$ to represent the angle between the three-momenta of particles labeled with four-momenta $i$ and $j$.

The maximum nuclear recoil will arise when
\bea
\theta_{pk'} = 0, \hspace{1cm} \theta_{\omega k'} = \pi, \hspace{1cm} \theta_{p\omega} = \pi.
\eea
Physically this is when the nucleus recoils in the same direction as the incident dark matter momentum, while the photon and scattered dark matter travel in a direction anti-parallel to the incoming dark matter. Including these relations leads to the equation
\begin{widetext}
\bea
E_R^2\left(\frac{m_T^2}{\mu^2}\right) + E_R\left(2\frac{m_T}{\mu}\left(\frac{\omega^2}{2m_\chi} + \omega(v+1)\right)-2m_T\left(\frac{\omega + m_\chi v}{m_\chi}\right)^2\right) + \left(\frac{\omega^2}{2m_\chi} + \omega(1 + v)\right)^2 = 0
\eea
\end{widetext}
Using the soft photon ($\omega/m_\chi \ll 1$ and $\omega \ll m_\chi v$) and non-relativistic ($v \ll 1$) limits, this reduces to
\be
E_R^2\left(\frac{m_T^2}{\mu_T^2}\right) + E_R\left(2m_T\frac{\omega}{\mu_T} - 2m_T v^2\right) + \omega^2 = 0.
\ee
This equation also arises when one takes the limit of phase space that describes the minimum recoil energy
\be
\theta_{pk'} = \pi, \hspace{1cm} \theta_{\omega k'} = 0, \hspace{1cm} \theta_{p\omega} = \pi.
\ee
The meaning of the quadratic equation is more evident if we define a rescaled recoil energy 
$x\equiv \left( \frac{m_T}{\mu_T} \right) E_R$.
The quadratic becomes simply %
\be
(x+\omega)^2= 2\mu_T v^2 x \,.
\ee
Two things are noteworthy in this result:
The first item is that the RHS breaks the $x \leftrightarrow \omega$ symmetry, so the $x$ and $\omega$ 
energy ranges are different.  The second item is that without the RHS,  $E_R$ or $\omega$ would be negative, 
and hence unphysical.  The RHS may fortunately be large, and so we do have a physical phase space. The solutions to the quadratic equation provide the phase space boundaries
\be
\label{eq:BremKin}
E_{R,\rm{max/min}} = \frac{\mu_T^2v^2}{m_T}\left[\left(1-\frac{\omega}{\mu_T v^2}\right) \pm \sqrt{1-\frac{2\omega}{\mu_T v^2}}\right].
\ee
These phase space limits are equivalent to those found for the Migdal effect in Eq.~(\ref{eq:MigdalKin}). Thus, we have the same global upper limit $\omega \le \frac{\mu_T v^2}{2} $.
Unsurprisingly, Eq.~(\ref{eq:MigdalKin}) is identical to Eq.~(\ref{eq:BremKin}) when $\cos\theta=1$ in Eq.~(\ref{eq:MigdalKin}) and with the (bremsstrahlung) photon energy $\omega$ substituted for the (Migdal) electromagnetic energy injection, $E_{\rm{EM}}$. Conceptually, the reason for this similarity is that here we are dealing with an inelastic collision between dark matter and the nucleus - the only difference from the case of the Migdal effect is that the excess energy goes into the production and emission of a photon with energy $\omega$. 

\subsection{Neutrino scattering and Migdal Effect}
\label{app:nuMigdalKinematics}
We start with the four momenta in the interaction $\nu(p) + A(k) \rightarrow \nu(p') + A(k')$, where $A$ is the atomic state and the four-momenta are
\bea
p &=& (E_\nu, \vec{p}),
\\
k &=& (m_T + E_{e,i},0),
\\
p' &=& (E_\nu', \vec{p'}),
\\
k' &=& (m_T +E_{e,f} + E_R, \vec{k'}).
\eea
Four momentum conservation requires
\bea
p - p' = k' - k.
\eea
We will use $p^2 = 0 = p'^2$ for relativistic neutrinos, $|\vec{k'}| = \sqrt{2m_TE_R}$, and the difference in atomic momenta is $k' - k = (E_R + \Delta E, \vec{k'})$ where $\Delta E = E_{e,f}-E_{e,i}$ is the sum of the outgoing unbounded electron energy, $E_{e}$ and energy from de-excitation, $E_{nl}$. Squaring both sides of the four-momentum equation we find
\bea
-2E_\nu E_\nu' (1-\cos\theta_{\nu\nu'}) = E_R^2 + \Delta E^2 + 2E_R \Delta E - 2m_T E_R.\nn
\eea
Substituting $E_\nu' = E_\nu - E_R - \Delta E$ and dropping the $E_R^2$ term we obtain the maximum recoil, which occurs at $\theta_{\nu\nu'} = \pi$, 
\bea
E_{R,{\rm{max}}} = \frac{(2E_\nu - \Delta E)^2}{2(m_T + 2E_\nu)}.
\eea
The minimum recoil occurs for $\theta_{\nu\nu'} \simeq 0$, leading to
\bea
E_{R,{\rm{min}}} = \frac{\Delta E^2}{2m_T}.
\eea
Note that this reproduces the limits found in \cite{Ibe:2017yqa}.

\subsection{Neutrino scattering and photon bremsstrahlung}
\label{app:nuBremKinematics}
For bremsstrahlung, the four-momenta in the lab frame can be written as
\bea
p &=& (E_\nu, \vec{p}),\\ 
k &=& (m_T, 0),\\
p' &=& (E_\nu',\vec{p}\,'),\\ 
k' &=& (m_T + E_R, m_T\vec{v}\,'_k),\\
\omega &=& (\omega,\vec{\omega}).
\eea
Energy conservation leads to
\bea
E_\nu = E_\nu' + E_R + \omega,
\eea
while three-momentum conservation gives
\bea
\vec{p} = \vec{p}\,' + m_T\vec{v}\,'_k + \vec{\omega}.
\eea
The maximum recoil energy is found when the recoiling nucleus is parallel to the incoming neutrino, and both the photon and scattered neutrino are anti-parallel to the incoming neutrino,
\bea
E_\nu = \sqrt{2m_T E_{R,\rm{max}}} - E_\nu' - \omega,
\eea
where we have used the non-relativistic form for $|\vec{k}'|$, neglecting terms of $\mathcal{O}(E_R^2)$. 
The minimum recoil energy is when the recoiling nucleus and photon are anti-parallel to the incoming neutrino, with the scattered neutrino traveling parallel to the incoming neutrino direction
\bea
E_\nu = -\sqrt{2m_T E_{R,\rm{min}}} + E_\nu' -\omega.
\eea
Combining the $E_{R,\rm{max}}$ equation with energy conservation, $E_\nu' = E_\nu - E_R - \omega$, leads to
\bea
-2E_{R,\rm{max}}(m_T + 2E_\nu) + 4E_\nu^2 = 0.
\eea
(Incidentally, this equation also holds to $\mathcal{O}(E_R^2)$, if one kept the next order term in solving for $|\vec{k}'|$ in terms of $E_R$). The solution to this equation is
\bea
E_{R,\rm{max}} = \frac{2E_{\nu}^2}{m_T + 2E_{\nu}}.
\eea
Next we can solve for $E_{R,\rm{min}}$ using the $E_{R,\rm{min}}$ equation and energy conservation. We find
\bea
E_{R,\rm{min}}(4\omega - 2m_T) + 4\omega^2 = 0.
\eea
Solving this equation, we find
\bea
E_{R,\rm{min}} = \frac{2\omega^2}{m_T - 2\omega}.
\eea

\bibliography{ddbrem}

\end{document}